\documentclass[12pt,a4paper]{article}
\usepackage{jheppub}  
\usepackage{amssymb} 
\usepackage{amsmath}
\usepackage{mathtools}
\usepackage{amsfonts}    
\usepackage{dsfont}
\usepackage{pdfpages}
\usepackage{verbatim}
\usepackage{tensor}
\usepackage{mathrsfs}
\usepackage[mathscr]{euscript}
\usepackage{tikz}
\usepackage{tensor}
\usetikzlibrary{calc}

\newcommand{\ba}{\begin{align}}

\newcommand{\be}{\begin{equation}}
\newcommand{\ee}{\end{equation}}
\def\bd{\begin{tikzpicture}}
\def\ed{\end{tikzpicture}}
\newcommand{\ket}[1]{| #1\rangle}
\newcommand*{\tran}{^{\mkern-1.5mu\mathsf{T}}}

\allowdisplaybreaks[1]

\title{$\boldsymbol{\text{AdS}_3/\text{CFT}_2}$ at higher genus}

\author{Lorenz Eberhardt} 
\affiliation{School of Natural Sciences, Institute for Advanced Study, \\
\hspace*{0.3cm}Princeton, NJ 08540, USA}
\emailAdd{elorenz@ias.edu}

\abstract{We continue our study of the worldsheet theory of superstrings on $\mathrm{AdS}_3 \times \mathrm{S}^3 \times \mathbb{T}^4$ in the tensionless limit \cite{Eberhardt:2019ywk}. We consider the theory on higher genus surfaces. We give evidence that the worldsheet correlators localise on certain worldsheets that cover the boundary of $\mathrm{AdS}_3$ holomorphically. This simplifies the string moduli space integral dramatically to a finite sum. This property shows that the higher genus corrections of the string worldsheet reproduce the structure of the $1/N$ corrections in the dual symmetric orbifold CFT $\mathrm{Sym}^N(\mathbb{T}^4)$.}

\begin{document}

\maketitle

\makeatletter
\g@addto@macro\bfseries{\boldmath}
\makeatother

\section{Introduction} \label{sec:intro}
The AdS/CFT correspondence \cite{Maldacena:1997re} has so far been mainly explored in its supergravity regime or at tree-level in string theory. 
In string theory, it is notoriously difficult to evaluate the moduli space integral at genera $\ge 2$, which hinders one to progress far in string perturbation theory. Of course, there have been various instances of simplified string theories in the literature, where this is not true, for example in the topological string \cite{Gopakumar:1998ki} or the minimal string \cite{Gross:1989vs, Douglas:1989ve}.

$\mathrm{AdS}_3$ offers a playground, where one can potentially go further, even for the physical string. The worldsheet theory is based on the $\mathrm{SL}(2,\mathds{R})$ WZW-model, which is conceivably fully solvable. This goal has only partially been achieved so far, see however \cite{Petropoulos:1989fc, Hwang:1990aq, Henningson:1991jc, Teschner:1997ft, Evans:1998qu, Teschner:1999ug, Maldacena:2000hw, Maldacena:2000kv, Maldacena:2001km, Teschner:2001gi} for important milestones in its understanding. String theory on $\mathrm{AdS}_3$ becomes simpler in the tensionless limit on $\mathrm{AdS}_3 \times \mathrm{S}^3 \times \mathbb{T}^4$. It was proposed in \cite{Gaberdiel:2018rqv, Eberhardt:2018ouy} that string theory for one unit of NS-NS flux becomes equivalent to the symmetric orbifold theory $\mathrm{Sym}^N(\mathds{T}^4)$. Thus, this proposal provides an example where both sides of the correspondence are in principle under control. In \cite{Gaberdiel:2018rqv, Eberhardt:2018ouy}, evidence for this proposal was given by computing the full tree-level string spectrum and matching it to the symmetric orbifold. It was found that contrary to the higher-flux case, the tensionless-point with one unit of NS-NS flux ($k=1$) does not have a long-string continuum in its spectrum and the only surviving world-sheet representations have $\mathfrak{sl}(2,\mathds{R})$ spin $j=\frac{1}{2}$.\footnote{We are referring here to the spin of the bosonic $\mathfrak{sl}(2,\mathds{R})_{k+2}$ algebra. The spin of the supersymmetric $\mathfrak{sl}(2,\mathds{R})_{k}$ algebra can be $0$, $\tfrac{1}{2}$ or $1$.}

These ideas were developed further in \cite{Eberhardt:2019ywk}, where genus 0 correlation functions were discussed. Evidence was given that the string theory moduli space integral for genus 0 \emph{localises} to a set of points, thus reducing the string theory correlation functions to finite sums over certain punctured Riemann spheres. This is exactly the structure that one find in the symmetric orbifold, where correlation functions can be reduced to correlators on certain covering spaces that were proposed to be identified with the worldsheet \cite{Pakman:2009zz, Eberhardt:2019ywk}. This mechanism makes the correspondence essentially manifest.\footnote{See also the series of works \cite{Gopakumar:2003ns, Gopakumar:2004qb, Gopakumar:2005fx, Aharony:2007fs}, where localisation in free gauge theories is discussed from a Feynman diagram point of view.}

It was furthermore argued in \cite{Eberhardt:2019ywk} that the value of the localised correlators follows \emph{exactly} from evaluating the on-shell action of the classical solution that corresponds to a given correlator. This makes a direct connection with the method of Lunin \& Mathur \cite{Lunin:2000yv, Lunin:2001pw} to evaluate the symmetric product correlators and shows abstractly their equivalence. In \cite{Dei:2019iym}, an alternative argument to show equivalence was put forward. It was shown that when the covering surface has genus 0, these correlators obey a differential equation and are hence fully fixed by symmetry up to an overall normalisation. 
\medskip

The goal of this paper is to establish that this picture persists for higher genus worldsheets. We will provide strong evidence that also in this case, the worldsheet correlators localise on punctured Riemann surfaces that are covering maps of a given punctured Riemann surface on which we want to compute the space-time correlator. 
We find that actually the $\mathrm{SL}(2,\mathds{R})_{k+2}$ WZW model possesses such a localising solution whenever we compute an $n$-point function in which the $\mathfrak{sl}(2,\mathds{R})$ spins satisfy
\be 
\sum_{i=1}^n j_i=\frac{k+2}{2}\big(n-2+2g\big)-\big(3g-3+n\big)\ , \label{eq:j constraint intro}
\ee
where $g$ is the genus of the Riemann surface. Remarkably, this condition is always true for the minimal flux case $k=1$, where only $j_i=\tfrac{1}{2}$ survives. We take the fact that this condition is always satisfied in the minimal flux case as evidence that this is the correct solution of the constraints on the correlator in this case, although we have not been able to give a full proof of this fact.

To our knowledge, the $\mathrm{SL}(2,\mathds{R})$ WZW model has never been studied at higher genus (apart from the thermal torus partition function). Thus, we spend some time discussing background on CFT on higher genus surfaces. We try to give a user-friendly introduction to these topics.
\medskip

This paper is organised as follows. In Section~\ref{sec:exact AdS3CFT2 correspondence}, we review some background and give a non-technical overview over the results. The reader only interested in the results and the ideas is invited to only read Section~\ref{sec:exact AdS3CFT2 correspondence}. In Section~\ref{sec:Ward identities}, we set up the technical machinery to compute correlators on higher Riemann surfaces. In Section~\ref{sec:constraint equations and recursion relations} we derive the constraints imposed by the affine symmetry on the spectrally flowed correlators.  We prove then in Section~\ref{sec: localising solution} that the localising solution given in \eqref{eq:localising solution} indeed solves these constraints. Since the constraints cannot be formulated in closed form, this proof is actually quite non-trivial. We end in Section~\ref{sec:discussion} with a discussion and future directions. For the benefit of the reader, we have included some relevant background material on Riemann surfaces and elliptic functions in Appendices~\ref{app:Riemann surfaces background} and \ref{app:elliptic functions}.

\section{An exact \texorpdfstring{$\text{AdS}_3/\text{CFT}_2$}{AdS3CFT2} correspondence} \label{sec:exact AdS3CFT2 correspondence}
In this section, we will give a motivation and a mostly non-technical overview of our results. Most of this was already explained in \cite{Eberhardt:2019ywk}.
\subsection{String theory on \texorpdfstring{$\text{AdS}_3 \times \text{S}^3 \times \mathbb{T}^4$}{AdS3xS3xT4} at \texorpdfstring{$k=1$}{k=1}}
String theory on $\text{AdS}_3 \times \text{S}^3 \times \mathbb{T}^4$ with $k$ units of NS-NS flux can be described in the RNS-formalism by the worldsheet theory \cite{Giveon:1998ns}
\be 
\mathfrak{sl}(2,\mathds{R})_k^{(1)} \oplus \mathfrak{su}(2)_k^{(1)} \oplus \mathds{T}^4 \oplus \text{ghosts}\ , \label{eq:RNS worldsheet theory}
\ee
where the superscript $(1)$ denotes an $\mathcal{N}=1$ superconformal algebra. We follow the conventions of \cite{Ferreira:2017pgt}. The $\mathrm{SL}(2,\mathds{R})$ WZW model based on the affine algebra $\mathfrak{sl}(2,\mathds{R})_k^{(1)} \cong \mathfrak{sl}(2,\mathds{R})_{k+2}\oplus \text{(3 free fermions)}$ features affine representations based on discrete representations and continuous representations of global $\mathfrak{sl}(2,\mathds{R})$, as well as their spectrally flowed versions. In terms of the $\mathfrak{sl}(2,\mathds{R})$ spin $j$, we have $j \in \mathds{R}$ for discrete representations and $j \in \tfrac{1}{2}+i \mathds{R}$ for continuous representations. The spin of discrete representations is moreover restricted to lie in the window $\big(\frac{1}{2},\frac{k+1}{2}\big)$ in order to be consistent with unitarity and modular invariance \cite{Henningson:1991jc, Evans:1998qu, Hwang:1990aq, Maldacena:2000hw, Eberhardt:2018vho}.

This picture changes drastically for $k=1$ \cite{Gaberdiel:2018rqv, Eberhardt:2018ouy}. Since $\mathfrak{su}(2)_1^{(1)} \cong \mathfrak{su}(2)_{-1}\oplus \text{(3 free fermions)}$, unitarity breaks down in the standard RNS-formalism. One way to make sense of the theory is to use the hybrid formalism of Berkovits, Vafa \& Witten \cite{Berkovits:1999im}, where the worldsheet theory is instead based on the supergroup WZW-model $\mathrm{PSU}(1,1|2)_k$, which remains well-defined even for $k=1$, but its representation content shrinks drastically. Intuitively, in terms of the RNS-formalism fields, only the representation $j=\tfrac{1}{2}$ survives on the worldsheet and the $\mathfrak{su}(2)_{-1}$ factor acts as ghosts, canceling four fermions and one boson on the worldsheet \cite{Gaberdiel:2018rqv}. Contrary to the $k>1$ case, the $k=1$ theory has a discrete spectrum. The worldsheet spectrum was evaluated in \cite{Eberhardt:2018ouy} and it was found that it matches precisely the symmetric orbifold of $\mathds{T}^4$ in the large $N$ limit.

\subsection{Equivalence of the worldsheet theory to the dual CFT}
The conjectured dual CFT to string theory on $\text{AdS}_3\times\text{S}^3\times\mathbb{T}^4$ for $k=1$ is the symmetric product orbifold of the sigma-model on $\mathbb{T}^4$. It is well-known how to compute correlation functions in this orbifold theory, see \cite{Arutyunov:1997gt, Arutyunov:1997gi,Jevicki:1998bm, Lunin:2000yv, Lunin:2001pw, Pakman:2009zz, Pakman:2009ab, Roumpedakis:2018tdb, Dei:2019iym} for computations using various different techniques. There is one common theme to all these methods. A correlator
\be 
\left\langle \mathcal{O}^{(w_1)}(x_1)\cdots  \mathcal{O}^{(w_n)}(x_n)\right\rangle \label{eq:dual CFT correlator}
\ee
of fields $\mathcal{O}^{(w_i)}(x_i)$ in the twisted sector $w_i$ receives contributions from all possible covering surfaces of the $n$-punctured Riemann sphere $\mathds{CP}^1\setminus\{x_1,\dots,x_n\}$ with given ramification indices $w_i$. 
For example, when computing a four-point function of fields in the 2-twisted sector $\mathcal{O}^{(2)}(x_i)$, there are five possible covering surfaces -- four distinct 4-punctured Riemann spheres and one 4-punctured torus, the latter being the pillow geometry \cite{Maldacena:2015iua}.
The problem of computing correlators in the orbifold theory can be reduced to computing correlation functions on these covering surfaces in the un-orbifolded theory. 

Lunin and Mathur \cite{Lunin:2000yv} used this idea to compute correlation functions of twist fields. The correlator on the Riemann sphere is lifted up to a correlator on the covering surface. This conformal transformation introduces a factor that can be written as a Liouville action evaluated on the Weyl factor $\phi$ of the conformal transformation,\footnote{We suppress various subtleties that appear for fermionic correlators. See \cite{Lunin:2001pw} for details.} see e.q.~\cite[Part 3, Chapter 2]{Deligne:1999qp} for a discussion
\begin{align}
\left\langle \mathcal{O}^{(w_1)}(x_1)\cdots  \mathcal{O}^{(w_n)}(x_n)\right\rangle=\sum_\Gamma \mathrm{e}^{-S_\text{L}[\phi]} \prod_{i=1}^n \partial \Gamma(z_i)^{-h_i}\bar{\partial} \bar{\Gamma}(z_i)^{-\bar{h}_i} \big\langle \widetilde{\mathcal{O}}(z_1)\cdots  \widetilde{\mathcal{O}}(z_n)\big\rangle_{\Sigma_{g,n}^\Gamma}\ , \label{eq:dual CFT correlator lift covering surface}
\end{align}
where $\widetilde{\mathcal{O}}(z_i)$ are the corresponding fields on the covering surface (that we have assumed to be primary). $z_i$ are the punctures of the covering surface $\Sigma_{g,n}^\Gamma$ that is selected out by the covering map. 
This expression is singular and has to be carefully regularised. This process is described in detail in \cite{Lunin:2000yv, Lunin:2001pw}.

It was suggested in \cite{Lunin:2000yv, Pakman:2009zz, Eberhardt:2019ywk} that these covering surfaces should be identified with the string worldsheet in the $\mathrm{AdS}_3/\mathrm{CFT}_2$ correspondence. At first glance, this seems wrong -- after all, in string theory one should integrate over all possible worldsheet geometries and not sum over discrete number of them. The way out is that the string moduli space integral should localise to these configurations. This would make the correspondence essentially manifest.

Such a localisation property on the worldsheet is a rather unusual behaviour for a worldsheet CFT. While localisation happens in flat space in the high-energy limit \cite{Gross:1987kza}, we expect the localisation to be an \emph{exact} behaviour of the CFT. 

We give strong arguments in this paper that such a localisation does indeed occur for $k=1$. We analyse correlation functions of spectrally flowed affine primaries of the $\mathrm{SL}(2,\mathds{R})_{k+2}$ WZW model on the worldsheet\footnote{We will get around using the hybrid formalism. In fact, the only needed input from the hybrid formalism is that only $j=\tfrac{1}{2}$ representations survive on the worldsheet.}
\be 
\left\langle V^{w_1}_{h_1}(x_1,z_1) \cdots  V^{w_n}_{h_n}(x_n,z_n)\right\rangle\ ,\label{eq:intro sphere correlator}
\ee
where the spectral flow $w_i$ is mapped under the correspondence to the twist $w_i$ in the corresponding correlator in the dual CFT \eqref{eq:dual CFT correlator} \cite{Giveon:2005mi, Giribet:2018ada}. Here, $z_i$ is the worldsheet position of the vertex operator and $x_i$ is a variable that transforms via M\"obius transformations under the global $\mathrm{SL}(2,\mathds{R})$ symmetry -- it can be regarded as the insertion point at the boundary of $\text{AdS}_3$. Finally, $h_i$ is the conformal weight under the global $\mathrm{SL}(2,\mathds{R})$.

It turns out that affine symmetry imposes certain recursion relations on these correlators that relates correlators with different values of $h_i$. They take the schematic form
\begin{multline}
\clubsuit \left\langle V^{w_i}_{h_i-1}(x_i,z_i) \prod_{j\ne i}^n  V^{w_j}_{h_j}(x_j,z_j)\right\rangle=\sum_{l=1}^n \spadesuit_l \left\langle V^{w_l}_{h_l+1}(x_l,z_l) \prod_{j\ne i}^n  V^{w_j}_{h_j}(x_j,z_j)\right\rangle\\
+\heartsuit\left\langle \prod_{j=1}^n  V^{w_j}_{h_j}(x_j,z_j)\right\rangle+\sum_{\mu=1}^g \left\langle \left(\diamondsuit^+_\mu J^+_{0,\mu}+\diamondsuit^3_\mu J^3_{0,\mu}+\diamondsuit^-_\mu J^-_{0,\mu}\right)\prod_{j=1}^n  V^{w_j}_{h_j}(x_j,z_j)\right\rangle\ . \label{eq:recursion relation schematic form}
\end{multline}
Here, the symbols $\clubsuit$, $\spadesuit_l$, $\heartsuit$ and $\diamondsuit_\mu^a$ are very complicated functions of the involved variables and act as differential operators in $x$-space. The zero-mode insertions $J_{0,\mu}^a$ introduce twisted boundary conditions along the $g$ $b$-cycles of the Riemann surface. Actually, we do not know a general closed form expression for the symbols $\clubsuit$, $\spadesuit_l$, $\heartsuit$ and $\diamondsuit_\mu^a$.

Rather remarkably, it turns out that these recursion relations simplify dramatically if one assumes that the correlation function localises on punctured Riemann surfaces that possess a covering map to the $n$-punctured Riemann sphere. In this case, the symbols $\clubsuit$, $\spadesuit_l$ and $\heartsuit$ become simple functions determined entirely by the corresponding covering map and $\diamondsuit_\mu^a=0$. These simplified recursion relations possess then a very simple solution which is given by\footnote{This is true provided that the constraint \eqref{eq:j constraint intro} is satisfied, which is always the case for $k=1$, see the discussion below eq.~\eqref{eq:j constraint intro}.}
\begin{multline} 
\left\langle \prod_{j=1}^n V_{h_j}^{w_j}(x_j,z_j)\right \rangle_{\!\!\Sigma_{g,n}}\\
=\sum_\Gamma  \delta^{(3g-3+n)}(\Sigma_{g,n}=\Sigma_{g,n}^\Gamma)  W_\Gamma(x_1,\dots,x_n) \prod_{i=1}^n (a^\Gamma_i)^{-h_i}(\bar{a}^\Gamma_i)^{-\bar{h}_i} \ . 
\end{multline}
Here, the right-hand side involves a sum over all possible covering maps $\Gamma$ that cover the given $n$-punctured Riemann sphere. To each covering map $\Gamma$, there is an associated covering surface $\Sigma^\Gamma_{g,n}$. The localisation of the solution now means that there is a $\delta$-function present that imposes the Riemann surface on which we are computing the correlator to be equal to the covering surface. The $h_i$ dependence of the answer is fully fixed by the recursion relations and takes a simple factorised form. Here, $a_i^\Gamma$ is a certain coefficient associated to the covering map, see eq.~\eqref{eq:Gamma Taylor expansion}. The remaining part of the correlator is not constrained by the recursion relations. 
It could be further constrained by using mixed Ward-identities, i.e.~the Knizhnik-Zamolodchikov equation \cite{Knizhnik:1984nr} in this spectrally flowed incarnation. However, the result is expected to be quite complicated for a higher genus surface and we have not tried to evaluate these constraints.

If one considers a string theory correlator, one would combine the vertex operator $V_{h_i}^{w_i}(x_i,z_i)$ with some internal vertex operator $V^\text{int}_i(z_i)$ of the sigma-model on $\mathbb{T}^4$ and integrate over moduli space. This leads to the genus-$g$ string correlator
\begin{multline}
\left\langle \prod_{j=1}^n V_{h_j}^{w_j}(x_j,z_j)\right \rangle_{\!\!\text{string}}\\
=g_\text{string}^{2g-2+n}\sum_\Gamma W_\Gamma(x_1,\dots,x_n) \prod_{i=1}^n (a^\Gamma_i)^{-h_i}(\bar{a}^\Gamma_i)^{-\bar{h}_i}  \left\langle \prod_{j=1}^n V^{\text{int}}_i(z_j^\Gamma)\right\rangle_{\Sigma_{g,n}^\Gamma}\ . \label{eq:string correlator}
\end{multline}
Here, $z_j^\Gamma$ are the punctures on the covering surface, that are determined by $\Gamma$. Thus, the string correlator reduces to the internal correlator evaluated on the covering surface, decorated with some universal prefactor that is independent of our choice of the internal vertex operators. Comparing this expression with \eqref{eq:dual CFT correlator lift covering surface}, we see that it has exactly the same structure. In fact, after proper regularisation of \eqref{eq:dual CFT correlator lift covering surface}, we see that the $h_i$ dependence of the prefactor is precisely the same.

\subsection{On-shell action}
What remains to be shown is that the rest of the prefactor $W_\Gamma(x_1,\dots,x_n)$ coincides with the regularised Liouville action. In the case of genus $0$, a semiclassical argument for this was given in \cite{Eberhardt:2019ywk}, which can be extended to the higher genus case as follows.

One starts by writing the semiclassical action of the $\mathrm{SL}(2,\mathds{R})$ sigma-model \cite{Giveon:1998ns, deBoer:1998gyt}\footnote{Since fermions are classically zero, we disregard them here. Moreover, we focus on $\mathrm{AdS}_3$, since we assumed the operators to sit in the vacuum of the internal CFT.}
\be\label{eq:AdS3action}
S_{\mathrm{AdS}_3}= \frac{k}{4\pi} \int \mathrm{d}^2 z\, \sqrt{g}\,  \Bigl( 4\,  \partial \Phi \, \bar\partial \Phi +\beta\,\bar{\partial}\gamma+\bar{\beta}\,\partial \bar{\gamma}- \mathrm{e}^{-2\Phi} \beta\bar{\beta} -k^{-1} R\, \Phi \Bigr)\ .
\ee
Here, $\gamma$ is a complex field that describes the boundary direction of $\mathrm{AdS}_3$. $\Phi$ is a real field parametrising the radial coordinate. The boundary of $\mathrm{AdS}_3$ is located at $\Phi \to \infty$. The complex field $\beta$ is auxiliary and could be integrated out again. The linear dilaton term is generated through renormalisation at the quantum level. It makes $\Phi$ transform anomalously under conformal transformations. The currents can be recovered easily from these fields via the Wakimoto construction \cite{Giveon:1998ns, deBoer:1998gyt}.
The classical solution describing the sphere correlator \eqref{eq:intro sphere correlator} is given by \cite{Eberhardt:2019ywk}
\begin{subequations}
\begin{align} 
\gamma(z)&=\Gamma(z)\ ,\\
\Phi(z,\bar{z})&=\sum_{i=1}^n \left(\frac{j_i}{k}-\frac{w_i}{2}\right)\log |z-z_i|^2+\sum_{a=1}^N \log |z-z_a^*|^2+\text{const.}\ ,
\end{align}
\end{subequations}
where $z_a^*$, $a=1,\dots,N$ are the $N$ poles of the covering map $\Gamma$. The solution for $\beta$ is more complicated and we shall not need it. The constant in $\Phi(z,\bar{z})$ is infinitely large and the worldsheet is hence `glued' to the boundary of $\mathrm{AdS}_3$. 

This solution generalises to the higher genus case as follows. One still has $\gamma(z)=\Gamma(z)$. For $\Phi(z,\bar{z})$, it is simpler to write down its derivative. $\partial \Phi(z)$ is holomorphic and is specified by the following conditions:
\begin{subequations}
\begin{align}
\mathop{\text{Res}}_{z=z_i} \partial \Phi(z)&=\frac{j_i}{k}-\frac{w_i}{2}\ , \\
\mathop{\text{Res}}_{z=z_a^*} \partial \Phi(z)&=1\ , \\
\int_{\alpha_\mu}  \partial \Phi(z)&=0\ , \qquad \mu=1,\dots,g\ .
\end{align}
\end{subequations}
Here, $\alpha_\mu$, $\mu=1,\dots,g$ are the $g$ $\alpha$-cycles of the Riemann surface, see Appendix~\ref{app:Riemann surfaces background}.
Moreover, $\partial \Phi(z)$ transforms under coordinate transformations as follows. For a conformal transformation $f$, we have
\be 
(f \cdot \partial \Phi)(z)=\frac{\Phi(f^{-1}(z))}{\partial f(f^{-1}(z))}+\frac{\partial^2 f(f^{-1}(z))}{2k(\partial f(f^{-1}(z)))^2}\ .
\ee
Because of its anomalous transformation behaviour, $\partial\Phi(z)$ obeys a modified residue theorem, which states that on the genus $g$ surface, we have
\be 
\sum_{p\in \Sigma_g} \mathop{\text{Res}}_{z=p} \partial \Phi(z)=\frac{1}{2k}\chi(\Sigma_g)=\frac{1-g}{k}\ .
\ee
Evaluating this leads precisely to the condition on $j$, see eq.~\eqref{eq:j constraint intro}. Clearly, the residue conditions determine $\partial \Phi(z)$ up to a holomorphic one-form, which in turn is fixed by the condition that the integral along the $\alpha_\mu$ cycles vanishes.

Let us now further restrict to the `ground state' solution, which describes the twisted sector ground state of the dual symmetric product orbifold. As shown in \cite{Dei:2019osr}, it is described by $j=\frac{k}{2}$. However, one notices that for this value of $j$, the $j$-constraint \eqref{eq:j constraint intro} is only satisfied when either $g=1$ or $k=1$.\footnote{In the case of the sphere, one can also insert one field at infinity, which transforms under the conjugate representation $j=1-\frac{k}{2}$. The $j$-constraint \eqref{eq:j constraint intro} can then also be satisfied for $g=0$ and arbitrary $k$.} So in the following we will further restrict to $k=1$, which is our main interest. In this case, we realise that the solution is given by
\be 
\partial \Phi(z)=-\frac{\partial^2 \Gamma(z)}{2\, \partial \Gamma(z)}\ ,
\ee
which has the correct residues and for $k=1$ also the correct transformation behaviour. Hence we conclude that
\be 
\Phi(z,\bar{z})=-\frac{1}{2}\log |\partial \Gamma(z)|^2+\text{const.}
\ee
and so $\phi(z,\bar{z})=-2\Phi(z,\bar{z})+\text{const.}$ can be identified with the Weyl factor of the conformal transformation given by $\Gamma(z)$. The additive constant is again formally infinite and hence the fields are `glued' to the boundary of $\mathrm{AdS}_3$. 
It was furthermore argued in \cite{deBoer:1998gyt} that the semiclassical on-shell is \emph{exact}, since the action becomes quadratic. The on-shell action coincides with the Liouville action for the field $\phi=-2\Phi+\text{const.}$ and we have
\be 
S_{\text{AdS}_3}[\Phi,\gamma,\beta]=S_\text{L}[\phi]=\frac{1}{8\pi} \int \mathrm{d}^2 z\, \sqrt{g}\,  \Bigl( 2\,  \partial \phi \, \bar\partial \phi  + R\, \phi \Bigr)+\text{const.}\ ,
\ee
where the right hand side is the Liouville action that appears in the conformal transformation of primary fields in a $c=6$ CFT, see \cite[eq.~(13.2)]{Friedan:1982is}. The constant arises from the constant shift in $\phi$, that is visible in the linear dilaton term.

Computing the correlator via the on-shell action hence leads precisely to the prefactor $\mathrm{e}^{-S_\text{L}[\phi]}$ that appears in \eqref{eq:dual CFT correlator lift covering surface} and hence suggests that (for $k=1$ and $j_i=\tfrac{1}{2}$), the prefactor in \eqref{eq:string correlator} is given by
\be 
W_\Gamma(x_1,\dots,x_n)=\text{const.}\times \exp\left(-S_\text{L}\!\left[\phi=\log|\partial \Gamma|^2\right]\right)\ .
\ee
The overall constant can be reabsorbed into a redefinition of the string coupling constant and this argument does not fix it.
\medskip

Hence these arguments go a long way towards proving the equivalence of correlation functions and hence the equivalence of the two theories.

\section{The Ward identities on a higher genus surface} \label{sec:Ward identities}
\subsection{Notation and setup}
Let us fix our notation and conventions for the $\mathrm{SL}(2,\mathds{R})$ WZW-model. We follow the conventions of \cite{Eberhardt:2019ywk}, which we shall briefly recall here. Since we are mainly interested in superstring theory on $\mathrm{AdS}_3$, we shift the level of the WZW model by two units, i.e.~we are considering $\mathrm{SL}(2,\mathds{R})_{k+2}$. See also the discussion below eq.~\eqref{eq:RNS worldsheet theory}.
\subsubsection{The $\mathrm{SL}(2,\mathds{R})_{k+2}$ current algebra}
The three currents satisfy the following OPEs:
\begin{subequations}
\begin{align}
J^3(z)J^3(w) &\sim -\frac{k+2}{2(z-w)^2}\ , \\ 
J^3(z)J^\pm(w) &\sim \pm\frac{J^\pm(w)}{z-w}\ , \\ 
J^+(z)J^-(w) &\sim \frac{k+2}{(z-w)^2}-\frac{2J^3(w)}{z-w}\ . 
\end{align}\label{eq:sl2R current algebra}%
\end{subequations}
The zero-modes $J_0^a$ generate the global $\mathrm{SL}(2,\mathds{R})$ symmetry of the model. In particular, $J_0^+$ is identified with the translation operator in the dual CFT. It is hence convenient to introduce a coordinate not only for the worldsheet position $z$, but also for the global $\mathrm{SL}(2,\mathds{R})$ symmetry, that we shall call $x$. The $x$-coordinate transforms under the symmetry action via M\"obius transformations.

The algebra \eqref{eq:sl2R current algebra} has an outer automorphism of the form
\begin{subequations}
\begin{align}
\widetilde{J}^\pm(z)=z^{\mp w}J^\pm(z)\ , \\
\widetilde{J}^3(z)=J^3(z)+\frac{(k+2)w}{2z}\ .
\end{align} \label{eq:spectral flow}%
\end{subequations}
The $\mathrm{SL}(2,\mathds{R})$ WZW model features importantly \emph{spectrally flowed representations}, that is, representations that are not conformal highest weight representations, but obtained by composition with the spectral flow automorphism.
\subsubsection{Primary vertex operators}
Next, we discuss spectrally flowed affine primary vertex operators, which are obtained from the spectral flow automorphism \eqref{eq:spectral flow}. They are characterised by the following quantum numbers: the spectral flow $w$, as it appears in eq.~\eqref{eq:spectral flow}, the representation of the zero modes before spectral flow $j$ (i.e.~its $\mathfrak{sl}(2,\mathds{R})$ spin) and its conformal weight $h$ with respect to the global $\mathrm{SL}(2,\mathds{R})$ symmetry. Our convention for an $\mathfrak{sl}(2,\mathds{R})$ representation of spin $j$ is
\be 
J^3 \ket{m}=m\ket{m}\ , \qquad J^\pm \ket{m}=(m\pm j) \ket{m\pm 1}\ , \label{eq:sl2R representation}
\ee
for $\mathfrak{sl}(2,\mathds{R})$ generators $J^\pm$ and $J^3$. The Casimir of such a representation is given by
\be 
\mathcal{C}=-(J^3)^2+\tfrac{1}{2}(J^+J^-+J^-J^+)=-j(j-1)\ .
\ee
Hence representations with $j \leftrightarrow 1-j$ are equivalent, as long as $m\pm j \ne 0$ for all $m$ in the representation, in which case the representation \eqref{eq:sl2R representation} truncates or becomes indecomposable.

We will suppress $j$ in our notation and write the spectrally flowed affine vertex operators as $V_h^w(x,z)$.
 They have defining OPEs
\begin{subequations}
\begin{align}
J^+(z) V_h^w(x,\zeta)&=\sum_{p=1}^{w+1} \frac{(J^+_{p-1} V_h^w)(x,\zeta)}{(z-\zeta)^p}+\text{reg.}\ , \\
\big(J^3(z)-x J^+(z)\big) V_h^w(x,\zeta)&=\frac{h V_h^w(x,\zeta)}{z-\zeta}+\text{reg.}\ , \\
\big(J^-(z)-2x J^3(z)+x^2 J^+(z)\big)  V_h^w(x,\zeta)&=\mathcal{O}((z-\zeta)^{w-1})\ .
\end{align}\label{eq:J V OPE}%
\end{subequations}
The shifts on the left-hand-side of thes OPEs can be derived by conjugating with the translation $\mathrm{e}^{x J_0^+}$, see also \cite{Eberhardt:2019ywk}.
We have moreover
\begin{subequations}
\begin{align}
(J_0^+ V_h^w)(x,\zeta)&=\partial_x V_h(x,z)\ , \label{eq:J0p action}\\
(J_{\pm w}^+ V_h^w)(x,\zeta)&=\left(h-\tfrac{k+2}{2}w\pm j\right) V_{h\pm 1}^w(x,\zeta)\ ,\label{eq:Jwp action}
\end{align}
\end{subequations}
since $J_0^+$ is the translation generator. \eqref{eq:Jwp action} follows from  \eqref{eq:spectral flow} and \eqref{eq:sl2R representation}.  It is convenient to define the $\mathfrak{sl}(2,\mathds{R})$ generators acting on the field $V_{h_i}^{w_i}(x_i,z_i)$ as
\begin{subequations}
\begin{align}
\mathcal{D}_i^+&=-\partial_{x_i}\ , \\
\mathcal{D}_i^3&=-(h_i+x_i\partial_{x_i})\ , \\
\mathcal{D}_i^-&=-(2h_ix_i+x_i^2\partial_{x_i})\ .
\end{align}\label{eq:action zero modes}
\end{subequations}
\subsection{The global Ward identities}
Let us first discuss the global Ward identities satisfied by the correlation function. For this, consider the correlator
\be 
\left\langle J^a(z) \prod_{j=1}^n V_{h_j}^{w_j}(x_j,z_j) \right \rangle
\ee
for $a=+,-,3$.
It is a single-valued holomorphic one-form in $z$ and hence the sum of all its residues has to vanish. In the coordinate $z$, singularities occur when $z=z_i$ for $i=1,\dots,n$. We thus have
\be 
0=\sum_{i=1}^n \mathop{\text{Res}}_{z=z_i}  \left\langle J^a(z) \prod_{j=1}^n V_{h_j}^{w_j}(x_j,z_j) \right \rangle=-\sum_{i=1}^n \mathcal{D}_i^a  \left\langle  \prod_{j=1}^n V_{h_j}^{w_j}(x_j,z_j) \right \rangle\ ,
\ee
and hence correlators satisfy global Ward identities
\begin{subequations}
\begin{align}
0&=\sum_{i=1}^n \partial_{x_i}  \left\langle  \prod_{j=1}^n V_{h_j}^{w_j}(x_j,z_j) \right \rangle\ , \\
0&=\sum_{i=1}^n (x_i\partial_{x_i}+h_i)  \left\langle  \prod_{j=1}^n V_{h_j}^{w_j}(x_j,z_j) \right \rangle\ , \\
0&=\sum_{i=1}^n (x_i^2\partial_{x_i}+2h_i x_i)  \left\langle  \prod_{j=1}^n V_{h_j}^{w_j}(x_j,z_j) \right \rangle\  .
\end{align}\label{eq:global Ward identities}%
\end{subequations}
\subsection{The kernel \texorpdfstring{$\Delta(z,\zeta)$}{Delta(z,zeta)}}
To formulate the local Ward identities, we need a particular kernel $\Delta(z,\zeta)$ on the Riemann surface.\footnote{Strictly speaking, $\Delta(z,\zeta)$ is defined on the universal covering space, since $\Delta(z,\zeta)$ fails to be periodic around the cycles of the surface.} It is a function in $\zeta$ and a one-form in $z$. It satisfies the following properties:
\begin{enumerate}
\item \label{item:Delta property 1} Quasi-periodicity:
\begin{subequations}
\begin{align} 
\Delta(z,\alpha_\mu(\zeta))&=\Delta(z,\zeta)\ , \\
\Delta(z,\beta_\mu(\zeta))&=\Delta(z,\zeta)+2\pi i \omega_\mu(z)\ , \\
\Delta( \alpha_\mu(z),\zeta)&=\Delta(z,\zeta)\ , \\
\Delta( \beta_\mu(z), \zeta)&=\Delta(z,\zeta)-2\pi i \omega_\mu(z)\ ,
\end{align}\label{eq:Delta quasi periodicity}%
\end{subequations}
for $\mu=1,\dots,g$.
Here, $\alpha_\mu$ and $\beta_\mu$ is a canonical choice of homology basis, as described in Appendix~\ref{subapp:period matrix}. $\alpha_\mu(z)$ denotes the endpoint of the path specified by $\alpha_\mu$, i.e.~the point that is once transported around the cycle $\alpha_\mu$. $\omega_\mu$ are the corresponding $g$ holomorphic one-forms on the Riemann surface, normalised such that
\be 
\int_{\alpha_\mu}\omega_\nu=\delta_{\mu\nu}\ .
\ee
\item \label{item:Delta property 2} As a one-form in $z$, $\Delta(z,\zeta)$ has the following simple poles:
\be 
z=\zeta,\, Q_1,\dots,\, Q_{g-1}\ ,
\ee
where $Q_1,\dots,\, Q_{g-1}$ are $g-1$ fixed points on the Riemann surface that do not depend on $\zeta$. The residue at all these points is one.
As a function in $\zeta$, the only singularity occurs for $\zeta=z$.
\end{enumerate}
For the torus, this object is well-known (and unique) and is given by (provided we choose flat coordinates)
\be 
\Delta(z,\zeta)=\frac{\vartheta_1'(z-\zeta|\tau)}{\vartheta_1(z-\zeta|\tau)}\ , \label{eq:torus Delta}
\ee
where $\vartheta_1(z|\tau)$ is the Jacobi theta-function. A similar construction works also for higher genus surfaces, see Appendix~\ref{app:Riemann surfaces background} for details. In the following we only need the existence of this object.
\subsection{The local Ward identities}
Next, we derive local Ward identities on the correlators, which determines the correlator with current insertions in terms of correlators without current insertions. They take the form
\begin{align}
&\left \langle J^a(z) \prod_{j=1}^n V_{h_j}^{w_j}(x_j,z_j) \right \rangle=-\sum_{i=1}^n \Delta(z,z_i) \mathcal{D}_i^a\left \langle \prod_{j=1}^n V_{h_j}^{w_j}(x_j,z_j) \right \rangle\nonumber\\
&\qquad+\sum_{i=1}^n\sum_{\ell=1}^{w_i} \frac{x_i^{1-a}\partial_{z_i}^\ell\Delta(z,z_i)}{\ell !}\left \langle (J^+_{\ell}V_{h_i}^{w_i})(x_i,z_i)   \prod_{j\ne i}^n V_{h_j}^{w_j}(x_j,z_j) \right \rangle\nonumber\\
&\qquad+ \sum_{\mu=1}^g \omega_\mu(z)\left \langle J^a_{0,\mu} \prod_{j=1}^n V_{h_j}^{w_j}(x_j,z_j) \right\rangle\ . \label{eq:local Ward identities}
\end{align}
By the exponent '$1-a$', we mean $1-a=0$ for $a=+$, $1-a=1$ for $a=3$ and $1-a=2$ for $a=-$. We also used the action of the zero-modes \eqref{eq:action zero modes}.

To derive them, consider the object
\be 
\Delta(z,\zeta)\left \langle J^a(\zeta) \prod_{j=1}^n V_{h_j}^{w_j}(x_j,z_j) \right \rangle
\ee
for $a=+,-,3$.
It is by construction a one-form in $\zeta$, though it is not periodic around the cycles of the surface. We can apply the residue theorem to it in $\zeta$. It has poles for $\zeta=z$ and $\zeta=z_i$ for $i=1,\dots,n$ and hence the sum over residues is
\begin{align}
-\left \langle J^a(z) \prod_{j=1}^n V_{h_j}^{w_i}(x_j,z_j) \right \rangle+\sum_{j=1}^n\mathop{\text{Res}}_{\zeta=z_j} \Delta(z,\zeta)\left \langle J^a(\zeta) \prod_{j=1}^n V_{h_j}^{w_j}(x_j,z_j) \right \rangle\ .
\end{align}
By the residue theorem, this equals the integral around the boundary of the fundamental domain, which is
\be 
\mathcal{C}=\alpha_1\beta_1 \alpha_1^{-1} \beta_1^{-1}\cdots \alpha_g\beta_g \alpha_g^{-1} \beta_g^{-1}\ , \label{eq:cut contour}
\ee
where juxtaposition denotes concatenation of paths.
See also Figure~\ref{fig:cut surface}. 
The $\beta_\mu$ integrals cancel, since $\beta_\mu$ and $\beta_\mu^{-1}$ appear pairswise in \eqref{eq:cut contour} and the integrand is periodic along the $\alpha_\mu$-cycles. The $\alpha_\mu$ integrals only partially cancel, due to the quasi-periodicity \eqref{eq:Delta quasi periodicity}. Thus, we have
\be 
\frac{1}{2\pi i}\oint_\mathcal{C} \Delta(\zeta,z)\left \langle J^a(\zeta) \prod_{j=1}^n V_{h_j}^{w_j}(x_j,z_j) \right\rangle=-\sum_{\mu=1}^g \omega_\mu(z) \oint_{\alpha_\mu}\left \langle J^a(\zeta) \prod_{j=1}^n V_{h_j}^{w_j}(x_j,z_j) \right\rangle\ .
\ee
Let us define 
\be 
J^a_{0,\mu} \equiv \oint_{\alpha_\mu} J^a(\zeta)\ .
\ee
Then we have shown that
\begin{multline}
\left \langle J^a(z) \prod_{j=1}^n V_{h_j}^{w_j}(x_j,z_j) \right \rangle=\sum_{i=1}^n\mathop{\text{Res}}_{\zeta=z_i} \Delta(z,\zeta)\left \langle J^a(\zeta) \prod_{j=1}^n V_{h_j}^{w_j}(x_j,z_j) \right \rangle\\
+ \sum_{\mu=1}^g \omega_\mu(z)\left \langle J^a_{0,\mu} \prod_{j=1}^n V_{h_j}^{w_j}(x_j,z_j) \right\rangle\ . 
\end{multline}
The residue can be further evaluated as follows. Recalling the OPE \eqref{eq:J V OPE}, we have
\begin{subequations}
\begin{align} 
\mathop{\text{Res}}_{\zeta=z_i} \Delta(z,\zeta)&\left \langle J^+(\zeta) \prod_{j=1}^n V_{h_j}^{w_j}(x_j,z_j) \right \rangle= \Delta(z,z_i) \partial_{x_i}\left \langle  \prod_{j=1}^n V_{h_j}^{w_j}(x_j,z_j) \right \rangle\nonumber\\
&\qquad+\sum_{\ell=1}^{w_i} \frac{\partial_{z_i}^\ell\Delta(z,z_i)}{\ell !}\left \langle (J^+_{\ell}V_{h_i}^{w_i})(x_i,z_j)   \prod_{j\ne i}^n V_{h_j}^{w_j}(x_j,z_j) \right \rangle\ , \\
\mathop{\text{Res}}_{\zeta=z_i} \Delta(z,\zeta)&\left \langle J^3(\zeta) \prod_{j=1}^n V_{h_j}^{w_j}(x_j,z_j) \right \rangle=\Delta(z,z_i) (x_i\partial_{x_i}+h_i)\left \langle \prod_{j=1}^n V_{h_j}^{w_j}(x_j,z_j) \right \rangle\nonumber\\
&\qquad+\sum_{\ell=1}^{w_i} \frac{x_i\partial_{z_i}^\ell\Delta(z,z_i)}{\ell !}\left \langle (J^+_{\ell}V_{h_i}^{w_i})(x_i,z_i)   \prod_{j\ne i}^n V_{h_j}^{w_j}(x_j,z_j) \right \rangle\ , \\
\mathop{\text{Res}}_{\zeta=z_i} \Delta(z,\zeta)&\left \langle J^-(\zeta) \prod_{j=1}^n V_{h_j}^{w_j}(x_j,z_j) \right \rangle= \Delta(z,z_i) (x_i^2\partial_{x_i}+2h_ix_i)\left \langle \prod_{j=1}^n V_{h_j}^{w_j}(x_j,z_j) \right \rangle\nonumber\\
&\qquad+\sum_{\ell=1}^{w_i} \frac{x_i^2\partial_{z_i}^\ell\Delta(z,z_i)}{\ell !}\left \langle (J^+_{\ell}V_{h_i}^{w_i})(x_i,z_i)   \prod_{j\ne i}^n V_{h_j}^{w_j}(x_j,z_j) \right \rangle\ .
\end{align}\label{eq:evaluating residues in local Ward identities}%
\end{subequations}
We have separated out the action of the zero modes and used that positive modes of $J^3(\zeta)$ and $J^-(\zeta)$ annihilate the field. Combining the ingredients, we arrive at \eqref{eq:local Ward identities}.

We should note that the resulting expression is not obviously periodic along the $\beta_\mu$-cycles, but is only so thanks to the global Ward identities \eqref{eq:global Ward identities}. All terms in the local Ward identities are manifestly periodic in $z \mapsto\beta_\mu(z)$, except for the first line in the eqs.~\eqref{eq:evaluating residues in local Ward identities}. Thus, we have
\begin{multline}
\left \langle J^a(\beta_\mu(z)) \prod_{j=1}^n V_{h_j}^{w_j}(x_j,z_j) \right \rangle-\left \langle J^a(z) \prod_{j=1}^n V_{h_j}^{w_j}(x_j,z_j) \right \rangle\\
=2\pi i \omega_{\mu}(z) \sum_{i=1}^n \mathcal{D}_i^a \left \langle \prod_{j=1}^n V_{h_j}^{w_j}(x_j,z_j) \right \rangle=0\ .
\end{multline}
Similarly, the expression seems to have poles as $z\to Q_i$, where $Q_i$ is one of the additional poles of $\Delta(z,\zeta)$ in $z$, see property~\ref{item:Delta property 2} of $\Delta(z,\zeta)$ above. The same argument shows however that these poles cancel out thanks to the global Ward identities.

We also remark that once that the correlator with a $J^+$ insertion is known, then the correlator with any other current insertion can be obtained. This is intuitively so, since they all sit in the same $\mathrm{SL}(2,\mathds{C})$ representation. To see this, we analyse the Ward identity resulting from the correlator
\begin{align}
\left\langle J^a(\zeta) J^b(z) \prod_{j=1}^n V_{h_j}^{w_j}(x_j,z_j) \right\rangle\ .
\end{align}
This is a periodic one-form in $\zeta$ and hence the sum over residues has to vanish. This results in
\be 
\sum_{i=1}^n\mathcal{D}_i^a \left\langle J^b(z) \prod_{j=1}^n V_{h_j}^{w_j}(x_j,z_j) \right\rangle=\tensor{f}{^{ab}_c}\left\langle J^c(z) \prod_{j=1}^n V_{h_j}^{w_j}(x_j,z_j) \right\rangle\ , \label{eq:relation between current insertions}
\ee
where $\tensor{f}{^{ab}_c}$ are the structure constants of $\mathfrak{sl}(2,\mathds{R})$.

Thus, the local Ward identities relate the correlator with current insertions to the following quantities: the original correlator of spectrally flown affine primaries, the correlator with zero-mode insertions along the $g$ cycles and correlators with positive mode actions of $J^+(\zeta)$ inserted. We will see below how these new unknowns are determined.
\subsection{The zero modes}\label{subsec:zero modes}
Let us discuss the zero-modes that come from integrating the currents along the cycles of the Riemann surface that appear in the local Ward identities \eqref{eq:local Ward identities}. We can get a clue about their meaning by analysing their periodicity property along the $\beta_\mu$ cycle of $z_i$. We have
\begin{align}
0&=\left \langle J^a(z)V_{h_i}^{w_i}(x_i,\beta_\mu(z_i))  \prod_{j\ne i}^n V_{h_j}^{w_j}(x_j,z_j) \right \rangle-\left \langle J^a(z) \prod_{j=1}^n V_{h_j}^{w_j}(x_j,z_j) \right \rangle\\
&=
\sum_{\nu=1}^g \omega_\nu(z)\left(\left \langle J^a_{0,\nu}V_{h_i}^{w_i}(x_i,\beta_\mu(z_i)) \prod_{j\ne i}^n V_{h_j}^{w_j}(x_j,z_j) \right\rangle-\left \langle J^a_{0,\nu} \prod_{j=1}^n V_{h_j}^{w_j}(x_j,z_j) \right\rangle\right)\nonumber\\
&\qquad-2\pi i \omega_\mu(z) \mathcal{D}_i^a \left \langle \prod_{j=1}^n V_{h_j}^{w_j}(x_j,z_j) \right \rangle\ ,
\end{align}
where we used that the other terms in the Ward identities \eqref{eq:local Ward identities} are manifestly periodic.
Thus, we have
\begin{multline}
\left \langle J^a_{0,\nu}V_{h_i}^{w_i}(x_i,\beta_\mu(z_i)) \prod_{j\ne i}^n V_{h_j}^{w_j}(x_j,z_j) \right\rangle\\
=\left \langle J^a_{0,\nu} \prod_{j=1}^n V_{h_j}^{w_j}(x_j,z_j) \right\rangle+2\pi i \delta_{\mu\nu}\mathcal{D}_i^a \left \langle \prod_{j=1}^n V_{h_j}^{w_j}(x_j,z_j) \right \rangle\ .
\end{multline}
Hence, correlators with zero-mode insertions are no longer periodic, but have twisted boundary conditions along the $\beta_\mu$-cycles.

Following \cite{Bernard:1987df, Bernard:1988yv}, we introduce correlators with twisted boundary conditions satisfying the boundary conditions
\begin{multline}
\left\langle V_{h_i}^{w_i} (x_i,\beta_\mu(z_i))\prod_{j\ne i}^n V_{h_j}^{w_j} (x_j,z_j) \right \rangle_{\gamma_1,\dots,\gamma_g}\\
= \gamma_\mu'(\gamma_\mu^{-1}(x_i))^{-h_i}\left\langle V_{h_i}^{w_i} (\gamma_\mu^{-1}(x_i),z_i)\prod_{j\ne i}^n V_{h_j}^{w_j} (x_j,z_j) \right \rangle_{\gamma_1,\dots,\gamma_g}\ , \label{eq:twisted boundary conditions}
\end{multline}
i.e.~the correlators implement the action of the $\mathrm{SL}(2,\mathds{C})$ group element $\gamma_\mu$, when transporting the fields along the cycle $\beta_\mu$.
As in  \cite{Bernard:1987df, Bernard:1988yv}, one can now \emph{define} the correlators with zero-mode insertions as
\begin{align}
\left \langle J^a_{0,\mu} \prod_{j=1}^n V_{h_j}^{w_j}(x_j,z_j) \right\rangle\equiv 2\pi i \frac{\mathrm{d}}{\mathrm{d}t}\Bigg|_{t=0}\left\langle\prod_{j=1}^n V_{h_j}^{w_j} (x_j,z_j) \right \rangle_{\gamma_\mu=\exp(t J_0^a),\gamma_{\nu\ne \mu}=\mathds{1}}\ . \label{eq:definition zero mode insertion}
\end{align}
Thus, to get a complete set of local Ward identities, one needs in principle also make a proposal about the twisted correlators. We will actually get around this difficulty for the localising solution, as we shall see in Section~\ref{subsec:proof}. In this case, the zero modes contribution will eventually drop out. Thus, we will not discuss twisted correlators further in the following.

We should also note that the three zero-mode insertions are related, since upon taking contour integrals of \eqref{eq:relation between current insertions}, we have
\be 
\sum_{i=1}^n\mathcal{D}_i^a \left\langle J^b_{\mu,0} \prod_{j=1}^n V_{h_j}^{w_j}(x_j,z_j) \right\rangle=\tensor{f}{^{ab}_c}\left\langle J^c_{\mu,0} \prod_{j=1}^n V_{h_j}^{w_j}(x_j,z_j) \right\rangle\ . \label{eq:relation between zero mode insertions}
\ee
\section{The constraint equations and the recursion relations} \label{sec:constraint equations and recursion relations}
\subsection{The constraint equations} \label{subsec:constraint equations}
While we have discussed how in principle the zero mode insertions in \eqref{eq:local Ward identities} are determined, we still have the unknowns
\be 
F^i_\ell\equiv\left \langle (J^+_{\ell}V_{h_i}^{w_i})(x_i,z_i)   \prod_{j\ne i}^n V_{h_j}^{w_j}(x_i,z_j) \right \rangle
\ee
appearing in the local Ward identities. They can be determined as for the sphere \cite{Eberhardt:2019ywk}, which we shall now review. First, we notice that $F^i_{w_i}$ is in principle determined as follows. The mode $J^+_{w_i}$ is the zero-mode before spectral flow and its action on the field $V_{h_i}^{w_i}(x_i,z_i)$ is determined by \eqref{eq:Jwp action}. We hence have
\be 
F^i_{w_i}=\left(h_i-\tfrac{k+2}{2}w_i+j_i\right)\left\langle V_{h_i+1}^{w_i}(x_i,z_i)\prod_{j\ne i}^n V_{h_j}^{w_j}(x_i,z_j) \right \rangle\ . \label{eq:Fiwi}
\ee
To write more uniform formulas in the following, we will also use eq.~\eqref{eq:J0p action} to write 
\be 
F^i_{0}=\left \langle (J^+_0 V_{h_i}^{w_i})(x_i,z_i)   \prod_{j\ne i}^n V_{h_j}^{w_j}(x_i,z_j) \right \rangle=\partial_{x_i}\left\langle \prod_{j=1}^n V_{h_j}^{w_j}(x_i,z_j) \right \rangle\ .
\ee
The other $F^i_\ell$'s can be obtained as follows. Let us consider the correlator
\begin{multline} 
\left \langle \left(J^-(z)-2 x_m J^3(z)+x_m^2 J^+(z)\right) \prod_{j=1}^n V_{h_j}^{w_j}(x_i,z_j) \right \rangle\\
=2\sum_{i=1}^n\Delta(z,z_i) h_i(x_i-x_m)\left \langle \prod_{j=1}^n V_{h_j}^{w_j}(x_j,z_j) \right \rangle
+\sum_{i=1}^n \sum_{\ell=1}^{w_i} \frac{(x_i-x_m)^2\partial_{z_i}^\ell\Delta(z,z_i)}{\ell !}F_\ell^{w_i} \\
+\sum_{\mu=1}^g \omega_\mu(z)\left \langle \left(J^-_{0,\mu}-2 x_m J^3_{0,\mu}+x_m^2 J^+_{0,\mu}\right) \prod_{j=1}^n V_{h_j}^{w_j}(x_i,z_j) \right \rangle\label{eq:constraining correlator}
\end{multline}
for a fixed $m \in \{1,\dots,n\}$.
The right-hand side of this equation is regular as $z\to z_m$. However, the correlator should implement the more stringent condition imposed by the OPE \eqref{eq:J V OPE}, which demands that this quantity vanishes to order $\mathcal{O}((z-z_m)^{w_m-1})$. In other words, we should require that
\begin{align}
\partial_z^r\left \langle \left(J^-(z)-2 x_m J^3(z)+x_m^2 J^+(z)\right) \prod_{j=1}^n V_{h_j}^{w_j}(x_j,z_j) \right \rangle\Bigg|_{z=z_m}=0 \label{eq:constraint equations}
\end{align} 
for $r=0,\dots,w_m-2$. This yields $\sum_{m=1}^n (w_m-1)$ constraints on the correlator that can be solved for the $\sum_{m=1}^n (w_m-1)$ unknowns $F_\ell^{w_i}$ for $i=1,\dots,n$ and $\ell=1,\dots,w_i-1$. We will refer to \eqref{eq:constraint equations} as the constraint equations.

One might be worried that the constraint equations might not be solvable, because the linear system is not invertible. This can indeed happen for certain values of the spectral flow parameters $w_i$. The condition for the constraint equations to be solvable appears to be
\be 
w_i \le 1+\sum_{j \ne i} w_j \label{eq:constraint solvability condition}
\ee
for all $i=1,\dots,n$. We have checked this in \texttt{Mathematica} for various small examples, but do not have an analytic proof for it.
This is the same condition that was found for the sphere correlators in \cite{Maldacena:2001km} and discussed in \cite{Eberhardt:2019ywk}. Thus, if this inequality is violated, the correlator of affine primary fields is forced to vanish.
\subsection{The recursion relations}
At this point, the correlator with a current insertion is fully fixed in terms of (possibly twisted) correlators of primary fields. However, we want to use this construction to derive \emph{constraints} on the correlators of the primary fields. 

For this, we observe that we know also an alternative expression for the leading term in the correlator \eqref{eq:constraining correlator}. We have
\begin{align}
&\Bigg \langle \big( J^-(z)- 2 x_m J^3(z)+ x_m^2 J^+(z)\big) \prod_{j=1}^n V_{h_j}^{w_j}(x_i,z_j) \Bigg \rangle\nonumber\\
&\ \ =\left \langle (J^-_{-w_m} V_{h_m}^{w_m})(x_m,z_m)   \prod_{j\ne m}^n V_{h_j}^{w_j}(x_i,z_j) \right \rangle+\mathcal{O}((z-z_m)^{w_m})\\
&\ \ =\left(h_m-\tfrac{k+2}{2}w_m-j_m\right)\left \langle  V_{h_m-1}^{w_m}(x_m,z_m)   \prod_{j\ne m}^n V_{h_j}^{w_j}(x_i,z_j) \right \rangle+\mathcal{O}((z-z_m)^{w_m})\ .
\end{align}
On the other hand, by using the expression in the form of eq.~\eqref{eq:constraining correlator} with the unknowns removed by solving the constraint equations \eqref{eq:constraint equations}, we can relate the correlator $\left \langle  V_{h_m-1}^{w_m}(x_m,z_m)   \prod_{j\ne m}^n V_{h_j}^{w_j}(x_i,z_j) \right \rangle$ to other correlators of primary fields and correlators with zero-mode insertions. We will refer to these equations as the recursion relations (as there are $n$ of them, one for every choice of $m$). We gave their schematic form in \eqref{eq:recursion relation schematic form}.
Unfortunately, these recursion relations are very hard to write down in closed form, since one first has to solve the system of the constraint equations \eqref{eq:constraint equations}. For illustration, we write down the recursion relations explicitly in the simplest case of $w_1=w_2=w_3=w_4=1$ and $g=1$, where no constraint equations have to be solved. Using M\"obius invariance to set $x_1=0$, $x_2=1$ and $x_3=\infty$, they take the form
\begin{align}
&\big(h_1-\tfrac{k+2}{2}-j_1\big)\left\langle V_{h_1-1}^{1}(x_1,z_1) \prod_{i=2}^4 V_{h_i}^{1}(x_i,z_i)\right\rangle=\Big((-h_1+h_2+h_3-h_4)\Delta(z_1,z_4)\nonumber\\
&\qquad+(h_1-h_2-h_3+h_4(1-2x_4))\Delta(z_1,z_3)+2h_4x_4\Delta(z_1,z_4)\Big)\left\langle \prod_{i=1}^4 V_{h_i}^{1}(x_i,z_i)\right\rangle\nonumber\\
&\qquad+\big(j_2-\tfrac{k+2}{2}+h_2)\partial_{z_2} \Delta(z_1,z_2)\left\langle V_{h_2+1}^{1}(x_2,z_2)\prod_{i\ne 2}^4 V_{h_i}^{1}(x_i,z_i)\right\rangle\nonumber\\
&\qquad+\big(j_3-\tfrac{k+2}{2}+h_3)\partial_{z_3} \Delta(z_1,z_3)\left\langle V_{h_3+1}^{1}(x_3,z_3)\prod_{i\ne 3}^4 V_{h_i}^{1}(x_i,z_i)\right\rangle\nonumber\\
&\qquad+\big(j_4-\tfrac{k+2}{2}+h_4)x_4^2\partial_{z_4} \Delta(z_1,z_4)\left\langle V_{h_4+1}^{1}(x_4,z_4)\prod_{i\ne 4}^4 V_{h_i}^{1}(x_i,z_i)\right\rangle\nonumber\\
&\qquad+x_4\big(-\Delta(z_1,z_2)+(1-x_4)\Delta(z_1,z_3)+x_4\Delta(z_1,z_4)\big)\partial_{x_4}\left\langle \prod_{i=1}^4 V_{h_i}^{1}(x_i,z_i)\right\rangle\nonumber\\
&\qquad +\left\langle J^-_{1,0} \prod_{i=1}^4 V_{h_i}^{1}(x_i,z_i)\right\rangle\ ,
\end{align}
where we chose $m=1$. There is a similar equation for $m=2,\, 3$ and $4$.

We will later derive the recursion relations in closed form under the simplifying assumption that the correlator localises on certain points in the moduli space of Riemann surfaces, see eq.~\eqref{eq:localised recursion relations}.

\section{The localising solution} \label{sec: localising solution}
The main purpose of this section is to show that the recursion relations always admit the very simple solution
\be 
\left\langle \prod_{j=1}^n V_{h_j}^{w_j}(x_j,z_j)\right \rangle=\sum_\Gamma  W_\Gamma(x_1,\dots,x_n) \prod_{i=1}^n (a^\Gamma_i)^{-h_i} \prod_{I=1}^{3g-3+n} \delta(f_I^\Gamma)\ , \label{eq:localising solution}
\ee
provided that the quantum numbers satisfy the condition
\be 
\sum_{i=1}^n j_i=\frac{k+2}{2}(n-2+2g)-(3g-3+n)\ , \label{eq:j constraint}
\ee
to which we shall refer to as the $j$ constraint. This is the solution we already discussed in Section~\ref{sec:exact AdS3CFT2 correspondence} and whose existence was already anticipated in \cite{Eberhardt:2019ywk}.

Here, the sum runs over branched covering maps $\Gamma$ with ramification indices $w_i$ at $z_i$. Such a covering map only exists for a discrete subset of the moduli space of genus $g$ surfaces with $n$ punctures $\mathcal{M}_{g,n}$, which is implemented by the constraints $f_I^\Gamma$. The coefficients $a_i^\Gamma$ are certain coefficients associated to the covering map. The function $W_\Gamma(x_1,\dots,x_n)$ is not constrained by the recursion relations, except that it satisfies the global Ward identities
\begin{align}
\sum_{i=1}^n x_i^{m+1}\partial_{x_i} W_\Gamma(x_1,\dots,x_n)=0 
\end{align}
for $m=-1,0,1$.

In the following, we shall review the ingredients that go into this solution. We then prove that this solution indeed satisfies the recursion relations in Section~ \ref{subsec:proof}. 

We should note that computing the actual value of $W_\Gamma(x_1,\dots,x_n)$ is presumably very hard. This is already the case in rational WZW models, where one uses null-vectors to constrain the correlators \cite{Mathur:1988yx, Mathur:1988rx}. In principle, it is however fixed up to an overall constant through the KZ-equations.
\subsection{Covering maps} \label{subsec:covering maps}
\subsubsection{Definition} \label{subsubsec:definition}
Let us introduce (branched) covering maps. A covering map $\Gamma: \Sigma_g \to \mathds{CP}^1$ is a holomorphic map satisfying the conditions:
\begin{enumerate}
\item \label{item:covering map first property} $\Gamma(z)$ maps $z_i$ to $x_i$, i.e.~
\be 
\Gamma(z_i)=x_i\ .
\ee
\item \label{item:covering map second property} Moreover, the points $z_i$ are ramification points of order $w_i$ of the map $\Gamma(z)$, meaning that $\Gamma(z)$ has the Taylor expansion
\be 
\Gamma(z)=x_i+a_i^\Gamma(z-z_i)^{w_i}+\mathcal{O}\big((z-z_i)^{w_i+1}\big)\label{eq:Gamma Taylor expansion}
\ee
around $z_i$ for some coefficient $a_i^\Gamma$ and $\Gamma(z)$ has no other critical point.\footnote{Our definition is a special case of a more general definition, where several $x_i$'s are allowed to coincide. In this case, the ramification is described by a partition $\lambda_i$. Physically, colliding $x_i$'s would correspond to multi-string configurations. We only analyse single-string correlation functions, where all $x_i$'s are distinct. Our arguments go however through also for the more general definition.}
\end{enumerate}
The degree $N$ of $\Gamma$ (i.e.~the number of preimages of a generic point $x \in \mathds{CP}^1$) can be computed by the Riemann-Hurwitz formula:
\begin{align}
N=1-g+\frac{1}{2}\sum_{i=1}^n(w_i-1)\ . \label{eq:Riemann Hurwitz}
\end{align}
\subsubsection{Existence and Hurwitz space} \label{subsubsec:existence and Hurwitz space}
Let us compute the (virtual) dimension of the space of covering maps for fixed $\Sigma_g$ and insertion points $z_i$, i.e.~for a fixed punctured Riemann surface $\Sigma_{g,n}$.  The map has degree $N$ given by \eqref{eq:Riemann Hurwitz}. Thus it has in particular $N$ poles (assuming that $\infty$ does not coincide with one of the $x_i$'s). The space dimension of the space of meromorphic functions with $N$ poles at fixed location is given by the Riemann Roch theorem, which yields\footnote{This is strictly speaking the virtual (i.e.~`expected') dimension, since the Riemann Roch theorem also has a correction term which is the dimension of the space of holomorphic vector fields with $N$ prescribed zeros. Holomorphic vector fields only exist on the sphere and the torus and there are three resp.~one of them corresponding to the three M\"obius transformations and the translation, respectively. Thus the correction term vanishes for $g=0$ and $n \ge 3$ or $g=1$ and $n \ge 1$ or $g \ge 2$. This is satisfied for all cases of interest and hence we shall suppress this subtlety.}
\be 
\text{dimension for $N$ fixed poles}=N+1-g\ . \label{eq:dimension N fixed poles}
\ee
The poles of $\Gamma$ can however be at arbitrary locations, which introduces $N$ further parameters. Thus, we have\footnote{The same conclusion can be also reached directly by computing $\text{H}^0(\Gamma^* T\mathds{CP}^1)-\text{H}^1(\Gamma^* T\mathds{CP}^1)=2N+1-g$ with the Riemann-Roch theorem and noticing that $\text{H}^0(\Gamma^* T\mathds{CP}^1)$ parametrises infinitesimal deformations of the map.}
\be 
\text{dimension for $N$ arbitrary poles}=2N+1-g\ .
\ee
We then have to impose the Taylor expansion \eqref{eq:Gamma Taylor expansion} around every point $z_i$, which leads to $\sum_{i=1}^n w_i$ additional conditions on the map. This means that the space of covering maps has expected dimension
\be 
\text{dimension of covering maps for fixed $\Sigma_{g,n}$}=2N+1-g-\sum_{i=1}^n w_i=3-3g-n\ ,
\ee
where we used the Riemann Hurwitz formula \eqref{eq:Riemann Hurwitz} to simplify the result. Thus the expected dimension is generically negative, which means that for a fixed punctured Riemann surface $\Sigma_{g,n}$, such a covering map does generically \emph{not} exist. 

Observing that this expected dimension equals $-\mathrm{dim}(\mathcal{M}_{g,n})$, we can however instead of focussing on a fixed punctured Riemann surface consider the space of all $n$-punctured Riemann surfaces of genus $g$, which introduces in turn $3g-3+n$ new parameters and hence
\be 
\text{dimension of space of covering maps for varying $\Sigma_{g,n}$}\ =0\ ,
\ee
meaning that the condition for a covering map to exist selects a discrete subset 
\be 
\mathcal{H}_g(x_1,\dots,x_n; w_1,\dots,w_n) \subset \mathcal{M}_{g,n}\ ,
\ee
known in the mathematics literature as the \emph{Hurwitz} space.\footnote{Mathematicians view the Hurwitz space as an $n$-dimensional manifold $\mathcal{H}(w_1,\dots,w_n)$, where the points $x_i$ are left arbitrary. It is known that the Hurwitz space is an $n$-dimensional complex connected manifold.}
The cardinality\footnote{Conventionally, one counts covering maps with enhanced automorphism group with a weighting factor $1/\mathrm{Aut}(\Gamma)$. The automorphism group of a map is the group of automorphisms $\Phi$ of the covering surface such that $\Gamma \circ \Phi=\Gamma$.
 For instance, the example of $g=1$ and $w_1=w_2=w_3=w_4=2$ that we will discuss in Section~\ref{subsubsec:examples} has an automorphism group of order 2 and hence $h_1(2,2,2,2)=\tfrac{1}{2}$. Similarly, the example of $g=1$ and $w_1=w_2=w_3=3$ has an automorphism group of order $3$ and hence $h_1(3,3,3)=\tfrac{1}{3}$. The cardinality is to be understood in this sense.}
\be 
h_g(w_1,\dots,w_n)\equiv |\mathcal{H}_g(x_1,\dots,x_n; w_1,\dots,w_n)|
\ee
is known as the Hurwitz number and is independent of $x_i$. It gives the number of terms in the sum of the localising solution \eqref{eq:localising solution}. It is a classical problem in algebraic geometry to compute the Hurwitz numbers and they have a combinatorial description in terms of the representation theory of the symmetric group $S_N$.\footnote{This relation should not come as a surprise to the reader, since the connection can be understood to be given via the symmetric product orbifold. See e.g.~\cite{Dei:2019iym} for a discussion.}

While this dimension counting shows that it could be possible to find covering maps, there are inequalities on the ramification indices that have to be obeyed for them to actually exist. Since $N \ge \max_{i=1,\dots,n}(w_i)$, it follows from the Riemann-Hurwitz formula that
\be 
(w_i-1)\le \frac{1}{2}\sum_{j=1}^n (w_j-1)-g\ , \qquad \sum_{j=1}^n(w_j-1) \in 2\mathds{Z} \label{eq:conditions for covering map to exist}
\ee
have to be satisfied 
for $i=1,\dots,n$ in order for a covering map to exist. Note that these are stronger conditions than the condition for the constraint equations to be solvable \eqref{eq:constraint solvability condition}. 

For every choice of $\Gamma$ in the sum of \eqref{eq:localising solution}, the points $z_i$ as well as the moduli of the Riemann surface $\Sigma_g$ are fixed. We can hence write down $3g-3+n$ functions on the moduli space of Riemann surfaces,\footnote{These functions do not have to be well-defined globally, but only in the vicinity of the localisation point.} such that their zero set is the localisation locus of the covering map. We denote such functions by $f_I^\Gamma$ for $I=1,\dots,3g-3+n$. Different choices are possible, but the ambiguity can always be absorbed into the function $W_\Gamma(x_1,\dotsm,x_n)$ in \eqref{eq:localising solution}.
\subsubsection{\texorpdfstring{$\mathrm{SL}(2,\mathds{C})$}{SL(2,R) covariance} covariance}  \label{subsubsec:SL2R covariance}
The covering map is covariant under $\mathrm{SL}(2,\mathds{C})$ transformations, which means that for $\gamma$ a M\"obius transformation, $\gamma(\Gamma(z))$ is also a covering map on the same surface for the points $\gamma(x_i)$. 
Let us write momentarily $\Gamma(z;x_1,\dots,x_n)$ to emphasize the dependence of the covering map on the points $x_i$. Let us choose $\gamma(x)=x+\varepsilon x^{m+1}$ for $m=-1,0,1$ for infinitesimal $\varepsilon$.
Then this implies that
\begin{align} 
\Gamma(z;\gamma(x_1),\dots,\gamma(x_n))&=\Gamma(z;x_1,\dots,x_n)+ \varepsilon\,\sum_{i=1}^n x_i^{m+1} \partial_{x_i} \Gamma(z;x_1,\dots,x_n) \\
&\overset{!}{=}\Gamma(z;x_1,\dots,x_n)+\varepsilon\, \Gamma(z;x_1,\dots,x_n)^{m+1} \ .
\end{align}
and hence
\be 
\sum_{i=1}^n x_i^{m+1} \partial_{x_i} \Gamma(z)=\Gamma(z)^{m+1} \label{eq:Moebius covariance}
\ee
for $m=-1,0,1$. This expresses the M\"obius covariance of the covering map.

The existence of $\Gamma(z)$ is hence invariant under M\"obius transformations, i.e.
\be 
\mathcal{H}_g(\gamma(x_1),\dots,\gamma(x_n),w_1,\dots,w_n)=\mathcal{H}_g(x_1,\dots,x_n,w_1,\dots,w_n)\ .
\ee
This also means that the constraints $\delta(f_I^\Gamma)$ are invariant under M\"obius transformations, i.e.
\be 
\sum_{i=1}^n x_i^{m+1} \partial_{x_i} \delta(f_I^\Gamma)=0 \label{eq:constraints Moebius invariance}
\ee
for $m=-1,0,1$ and $I=1,\dots,3g-3+n$.
\subsubsection{The coefficients \texorpdfstring{$a_i^\Gamma$}{aiGamma}}  \label{subsubsec:ai}
The coefficients $a_i^\Gamma$ that appear in the Taylor expansion of the covering map \eqref{eq:Gamma Taylor expansion} play an important role in the story, since they feature in the localising solution \eqref{eq:localising solution}.

They are also $\mathrm{SL}(2,\mathds{C})$ covariant, which follows readily from expanding \eqref{eq:Moebius covariance} around the insertion points,
\begin{align}
\sum_{i=1}^n x_i^{m+1} \partial_{x_i} a_j^\Gamma =(m+1)x^m_j a_j^\Gamma\ .
\end{align}
These facts imply that the localising solution \eqref{eq:localising solution} satisfies the global Ward identities \eqref{eq:global Ward identities}.
\subsubsection{Examples} \label{subsubsec:examples}
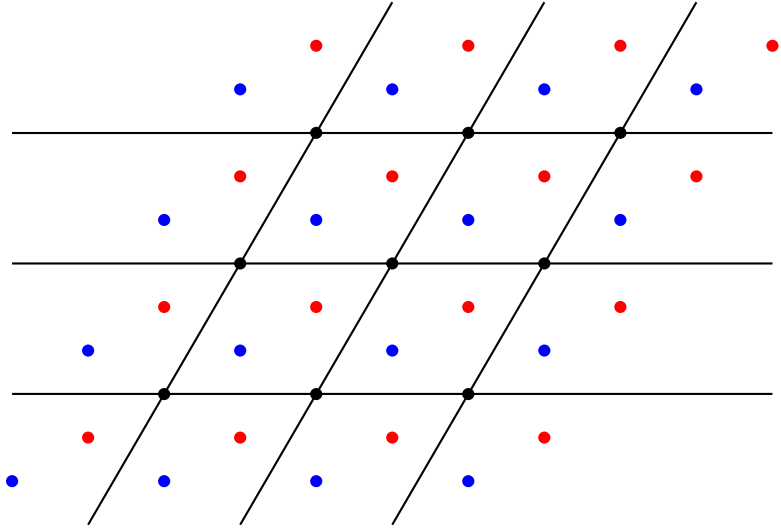
\begin{figure}
\begin{center}
\begin{tikzpicture}
\draw[thick] (180:5) -- (0:5);
\draw[thick] ($(180:5)+(90:1.73)$) -- ($(0:5)+(90:1.73)$);
\draw[thick] ($(180:5)+(270:1.73)$) -- ($(0:5)+(270:1.73)$);
\draw[thick] ($(240:4)$) -- ($(60:4)$);	
\draw[thick] ($(240:4)+(2,0)$) -- ($(60:4)+(2,0)$);	
\draw[thick] ($(240:4)+(-2,0)$) -- ($(60:4)+(-2,0)$);	
\foreach\m in {-1,0,1}{
	\foreach\n in {-1,0,1}{
		\fill[thick] (2*\m+\n,1.732*\n) circle (.08);
	}
}
\foreach\m in {-1,0,1,2}{
	\foreach\n in {-2,-1,0,1}{
		\fill[thick,red] (2*\m+\n,1.732*\n+1.155) circle (.08);
	}
}
\foreach\m in {-2,-1,0,1}{
	\foreach\n in {-1,0,1,2}{
		\fill[thick,blue] (2*\m+\n,1.732*\n-1.155) circle (.08);
	}
}
\end{tikzpicture}
\caption{The hexagonal lattice. The black points are $z_1=0$, the blue points are $z_2$ and the red points are $z_3$. One can clearly see that there is an order 3 automorphism that rotates the picture around the origin.} \label{fig:hexagonal lattice}
\end{center}
\end{figure}

\paragraph{$w_1=w_2=w_3=w_4=2$ and $g=1$.} Let us start with the classical example of covering the sphere by a torus, branched over four points with ramification indices 2. In this case, the Hurwitz number is $\tfrac{1}{2}$ and hence there will be one covering map with an automorphism group of order 2. The degree as computed from the Riemann-Hurwitz formula \eqref{eq:Riemann Hurwitz} is 2 and hence the covering map can be written as 
\be 
\Gamma(z)=\frac{A\wp(z|\tau)+B}{C\wp(z|\tau)+D}
\ee
for appropriate constants $A$, $B$, $C$ and $D$, since every degree 2 elliptic function can be written in this way. See Appendix~\ref{app:elliptic functions} for some background on elliptic functions that we shall use frequently. $z_i$ are fixed to be the critical points of $\Gamma(z)$. We have
\be 
\partial\Gamma(z)=\frac{(AD-BC) \wp'(z|\tau)}{(C\wp(z|\tau)+D)^2}=0 \quad \Longleftrightarrow\quad z=0,\,\tfrac{1}{2},\, \tfrac{\tau}{2},\,\tfrac{\tau+1}{2}\ ,
\ee
where we used the form of the zeros of the derivative of the Weierstrass $\wp$ function. Thus, we set $z_1=0$, $z_2=\tfrac{1}{2}$, $z_3=\tfrac{\tau}{2}$, $z_4=\tfrac{\tau+1}{2}$.
We determine then $A,B,C$ and $D$ such that $\Gamma(z_i)=x_i$ for $i=1,2,3$, which determines $\Gamma(z)$ to be
\begin{align}
\Gamma(z)=\frac{x_1(x_2-x_3)\wp(z|\tau)+x_3(x_1-x_2)\wp(\frac{1}{2}|\tau)+x_2(x_3-x_1)\wp (\frac{\tau}{2}|\tau)}{(x_2-x_3)\wp(z|\tau)+(x_1-x_2)\wp(\frac{1}{2}|\tau)+(x_3-x_1)\wp (\frac{\tau}{2}|\tau)}\ .
\end{align}
Finally, we have to satisfy the condition
\begin{align}
\frac{x_1(x_2-x_3)\wp(\frac{\tau+1}{2}|\tau)+x_3(x_1-x_2)\wp(\frac{1}{2}|\tau)+x_2(x_3-x_1)\wp (\frac{\tau}{2}|\tau)}{(x_2-x_3)\wp(\frac{\tau+1}{2}|\tau)+(x_1-x_2)\wp(\frac{1}{2}|\tau)+(x_3-x_1)\wp (\frac{\tau}{2}|\tau)}=x_4\ ,
\end{align}
so that $\Gamma(z_4)=x_4$. This determines $\tau$ uniquely in terms of the cross ratio and the solution can be written as, see e.g. \cite[Section 5.4.]{Zagier}
\be 
\tau=i \, \frac{\tensor[_2]{F}{_1}\big(\frac{1}{2},\frac{1}{2};1;\frac{(x_1-x_3)(x_4-x_2)}{(x_2-x_3)(x_4-x_1)}\big)}{\tensor[_2]{F}{_1}\big(\frac{1}{2},\frac{1}{2};1;\frac{(x_1-x_2)(x_3-x_4)}{(x_2-x_3)(x_4-x_1)}\big)}\ . \label{eq:2222 tau solution}
\ee 
This is the unique covering map. Since it is an even map, it has an automorphism group of order 2 generated by $z \mapsto -z$, which fixes $z_1$, \dots, $z_4$. The coefficients $a_i^\Gamma$ are straightforward to evaluate. They are given by
\begin{align}
a_1^\Gamma&=\frac{\pi^2(x_1-x_2)(x_3-x_1)}{x_2-x_3} \vartheta_3(\tau)^4\ , \\
a_2^\Gamma&=-\frac{\pi^2(x_1-x_2)(x_2-x_3)}{x_3-x_1} \vartheta_4(\tau)^4\ , \\
a_3^\Gamma&=-\frac{\pi^2(x_2-x_3)(x_3-x_1)}{x_1-x_2} \vartheta_2(\tau)^4\ , \\
a_4^\Gamma&=\frac{\pi^2(x_1-x_4)(x_2-x_3)}{x_3-x_4}\vartheta_4(\tau)^4\ ,
\end{align}
where $\tau$ is given by \eqref{eq:2222 tau solution} and $\vartheta_\nu(\tau)$ are the Jacobi theta functions.
\paragraph{$w_1=w_2=w_3=3$ and $g=1$.} Next, we consider a torus covering the sphere branched over three-points with ramification index 3 each. From the Riemann-Hurwitz formula, we see that such a map must have degree 3. It will turn out that the corresponding Hurwitz number is $\tfrac{1}{3}$, i.e.~there is a unique such map, with automorphism group of order 3. This fixes the torus to be the hexagonal torus, i.e.~$\tau=\mathrm{e}^{\frac{\pi i}{3}}$, since this the only torus with an automorphism group of order $\ge 3$. Also the insertion points of the covering surface have to be invariant under an order 3 automorphism, which fixes them to be
\be 
z_1=0\ , \qquad z_2=\frac{1}{3}+\frac{\tau}{3}\ , \qquad z_3=\frac{2}{3}+\frac{2\tau}{3}\ ,
\ee
with $\tau=\mathrm{e}^{\frac{\pi i}{3}}$. See Figure~\ref{fig:hexagonal lattice} for a schematic depiction of the corresponding lattice. Requiring the ramification around $z_1=0$ fixes the covering map to have the form
\be 
\Gamma(z)=\frac{A \wp'(z|\tau)+B}{C \wp'(z|\tau)+D}
\ee
for some constants $A,B,C$ and $D$. We have
\be 
g_2=0\ , \qquad \wp(z_2|\tau)=\wp(z_3|\tau)=0
\ee
for the hexagonal lattice. The latter follows from the duplication formula \eqref{eq:Weierstrass duplication formula}. 
From the differential equation of the Weierstrass $\wp$ function  \eqref{eq:Weierstrass differential equation} we then also conclude that\footnote{Incidentally, $g_3=(2\pi)^{-6}\Gamma(\tfrac{1}{3})^{18}$ for the hexagonal lattice \cite{Stiller}.}
\be 
\wp'(z_2|\tau)=i \sqrt{g_3}\ , \qquad  \wp'(z_3|\tau)=-i \sqrt{g_3}\ .
\ee
The derivatives of $\Gamma(z)$ are given by 
\begin{align}
\partial\Gamma(z)&= \frac{6(AD-BC)\wp(z|\tau)^2}{(C\wp'(z|\tau)+D)^2}\ , \\
\partial^2\Gamma(z)&= \frac{12(AD-BC)\wp(z|\tau)(-2\wp(z|\tau)^3-g_3+\wp'(z|\tau) D)}{(C\wp'(z|\tau)+D)^3}
\end{align}
and hence the first two derivatives at $z=z_2$ and $z=z_3$ indeed vanish, confirming that $z_2$ and $z_3$ are ramification points of order 3. $A,B,C$ and $D$ can again be determined such that $\Gamma(z_i)=x_i$ is true.
The coefficients $a_i^\Gamma$ are given by
\begin{subequations}
\begin{align}
a_1^\Gamma&=\frac{i\sqrt{g_3} (x_1-x_2)(x_1-x_3)}{x_2-x_3}\ , \\
a_2^\Gamma&=\frac{i\sqrt{g_3} (x_1-x_2)(x_2-x_3)}{x_1-x_3}\ , \\
a_3^\Gamma&=\frac{i\sqrt{g_3} (x_1-x_3)(x_2-x_3)}{x_1-x_2}\ .
\end{align}
\end{subequations}
\subsection{Non-renormalisation}
We now discuss some striking consequences of the conjectured form of the solution. They already partially appeared in \cite{Eberhardt:2019ywk}.

From \eqref{eq:Riemann Hurwitz}, it follows that there is a maximal genus that contributes to any given correlator. This is a rewriting of \eqref{eq:conditions for covering map to exist}
\be 
g\le \frac{1}{2} \left(\sum_{j \ne i} (w_j-1)-(w_i-1)\right)\ . \label{eq:genus bound}
\ee
Hence the string perturbation series \emph{truncates} in this case. This is very different from the usual string genus expansion, where the genus $g$ contribution diverges as $(2g)!$ \cite{Shenker:1990uf, Aniceto:2011nu}, indicating the presence of non-perturbative objects of mass $\sim g_\text{string}^{-1}$ -- D-branes. Thus the theory in flat space has a complicated non-perturbative completion. In the present case, the genus expansion truncates and hence a priori \emph{no} non-perturbative objects are needed to complete the theory.

This conclusion is even more stringent for two-point functions, i.e.~for the string spectrum. Since in this case we need $w_1=w_2$ for a covering map to exist,  eq.~\eqref{eq:genus bound} becomes $g\le 0$. Hence the string spectrum is \emph{tree-level exact}. The tree-level string spectrum of strings on $\text{AdS}_3 \times \text{S}^3 \times \mathbb{T}^4$ was computed in \cite{Eberhardt:2018ouy} and was found to match the symmetric orbifold $\text{Sym}^N(\mathbb{T}^4)$ in the limit $N \to \infty$. Our analysis implies that this should remain true to all orders in $1/N$. Since the perturbation series truncates, one would also expect the absence of non-perturbative corrections. Taken together, these results indicate the matching of the exact spectrum (in $\alpha'$ and $g_\text{string}$) in this instance of the $\text{AdS}_3/\text{CFT}_2$ correspondence.
\subsection{The proof}\label{subsec:proof}
Now we shall present the proof that the ansatz \eqref{eq:localising solution} indeed solves the recursion relations described in Section~\ref{sec:constraint equations and recursion relations}, provided that the constraint \eqref{eq:j constraint} on the spins is satisfied. This will provide strong evidence for its correctness. We will divide the proof in several small steps.

It will be convenient to choose a map $\Gamma(z)\equiv  \Gamma(z_1,\dots,z_n,x_1,\dots,x_n,\boldsymbol{\Omega};z)$ that defines an extension of the covering map, i.e. we require that when the localisation constraints are satisfied, $\Gamma(z)$ equals the covering map, but it is otherwise arbitrary.  We even allow $\Gamma(z)$ to be non-periodic (and hence only defined on the universal covering map of $\Sigma_g$). Here, $\boldsymbol{\Omega}$ is the period matrix of the surface and represents the $3g-3$ moduli of $\Sigma_g$.

Let us define
\begin{align}
G(z)&\equiv\left\langle\left( J^-(z)-2\Gamma(z) J^3(z)+\Gamma(z)^2 J^+(z)\right) \prod_{j=1}^n V_{h_j}^{w_j}(x_j,z_j) \right \rangle\ ,\label{eq:G definition} \\
H(z)&\equiv\left\langle\left(J^3(z)-\Gamma(z) J^+(z)\right) \prod_{j=1}^n V_{h_j}^{w_j}(x_j,z_j) \right \rangle\ .\label{eq:H definition}
\end{align}
Note that these definitions make sense thanks to the extension of the covering map we have chosen above. However they a priori depend on the choice of $\Gamma(z)$.

We analyse in the following the one-forms $G(z)$ and $H(z)$ for the localised solution \eqref{eq:localising solution}. Obviously, also $G(z)$ and $H(z)$ are then localised. We should note that $G(z)$ and $H(z)$ contain a priori terms that are proportional to a product of delta functions of the form $\prod_{I=1}^{3g-3+n} \delta(f_I)$, but also derivatives of $\delta$ functions of the form $\delta'(f_I)\prod_{J\ne I}^{3g-3+n} \delta(f_J)$ appear. These terms are present, since the input parameters of the constraint equations  (namely the residues of the first order poles $\partial_{x_i} \left\langle \prod_{j=1}^n V_{h_j}^{w_j}(x_j,z_j)\right\rangle$ and the zero mode insertions $\left\langle J^a_{0,\mu} \prod_{j=1}^n V_{h_j}^{w_j}(x_j,z_j)\right\rangle$) possess them. Note that because of the distributional identity
\be 
x \delta'(x)=-\delta(x)\ , \label{eq:delta distributional identity}
\ee
we can extract a term proportional to $\delta'(f_I)\prod_{J\ne I}^{3g-3+n} \delta(f_J)$ in $G(z)$ or $H(z)$ by multiplying with the constraint $f_I$.
\subsubsection{Step 1: \texorpdfstring{$G(z)$}{G(z)} has no derivatives of \texorpdfstring{$\delta$}{delta} functions}  \label{subsubsec:step 1}
We will first show that while the terms defining $G(z)$ involve in general derivatives of $\delta$ functions, these terms cancel out in $G(z)$. To show this, let us consider
\be 
G_I(z)=f_I G(z)\ ,
\ee
Thus, $G_I(z)$ contains by construction no derivatives of $\delta$ functions. Hence, we may use that $\Gamma(z)$ is only evaluated on its localisation locus. It follows that for $r=0,\dots,w_m-2$
\begin{align} 
\partial^r_z &G_I(z=z_m)\nonumber\\
&=f_I\partial_z^r\left \langle \left(J^-(z)-2 \Gamma(z) J^3(z)+\Gamma(z)^2 J^+(z)\right) \prod_{j=1}^n V_{h_j}^{w_j}(x_i,z_j) \right \rangle\Bigg|_{z=z_m}\\
&=f_I\partial_z^r\left \langle \left(J^-(z)-2 x_m J^3(z)+x_m^2 J^+(z)\right) \prod_{j=1}^n V_{h_j}^{w_j}(x_i,z_j) \right \rangle\Bigg|_{z=z_m}=0\ ,
\end{align}
where we used the constraint equations \eqref{eq:constraint equations}. The replacement $\Gamma(z) \mapsto x_m$ would only influence the correlator at order $(z-z_m)^{w_m-1}$ and hence does not change the result. Furthermore, $G_I$ can only have poles where $\Gamma(z)$ has a pole and it has in fact a double pole there. We thus see that
\be 
\frac{G_I(z)}{\partial \Gamma(z)}
\ee
is a holomorphic function that does not have any poles and is hence constant. It follows that
\be 
G_I(z)=A_I \partial \Gamma(z)\ .
\ee
We finally compute the proportionality constant $A_I$. Let $N$ be the degree of the covering map. Let us denote by $z_1^*,\dots,z_N^*$ the $N$ poles of the covering map. We have
\begin{align}
A_I N &=-\sum_{a=1}^N \mathop{\text{Res}}_{z=z_a^*} \frac{G_I(z)}{\Gamma(z)} \label{eq:AI line 1}\\
&=-\sum_{a=1}^N \mathop{\text{Res}}_{z=z_a^*}  f_I\Gamma(z)\left\langle J^+(z) \prod_{j=1}^n V_{h_j}^{w_j}(x_j,z_j) \right \rangle\label{eq:AI line 2}\\
&=\sum_{i=1}^n\mathop{\text{Res}}_{z=z_i}  f_I\Gamma(z)\left\langle J^+(z) \prod_{j=1}^n V_{h_j}^{w_j}(x_j,z_j) \right \rangle\label{eq:AI line 3}\\
&=\sum_{i=1}^n \mathop{\text{Res}}_{z=z_i}  f_I(x_i+a_i^\Gamma(z-z_i)^{w_i})\left\langle J^+(z) \prod_{j=1}^n V_{h_j}^{w_j}(x_j,z_j) \right \rangle\label{eq:AI line 4}\\
&=\sum_{j=1}^n f_I(x_j F_0^j+a_j^\Gamma F_{w_j}^j) \label{eq:AI line 5}\\
&=\sum_{j=1}^n f_I\left(-h_j\left\langle \prod_{i=1}^n V_{h_i}^{w_i}(x_i,z_i) \right \rangle +a_j^\Gamma F_{w_j}^j\right)=0\ . \label{eq:AI line 6}
\end{align}
We have used the following properties. In \eqref{eq:AI line 1}, we used that $\partial \Gamma(z)/\Gamma(z)$ has a pole with residue $-1$ at $z_a^*$. In \eqref{eq:AI line 2}, we inserted the definition of $G_I(z)$ and used the fact that only terms with a second order pole, i.e.~the last term in the definition \eqref{eq:G definition} may contribute to the residue. In \eqref{eq:AI line 3}, we used that the sum over all residues of the integrand vanishes and thus, we can alternatively sum over the other residues located at $z=z_i$.\footnote{The integrand is periodic along the cycles, since $f_I \Gamma(z)$ does not depend on the extension of the covering map.} In \eqref{eq:AI line 4}, we inserted the expansion of $\Gamma(z)$ \eqref{eq:Gamma Taylor expansion} (which is possible because of the presence of the localisation constraint $f_I$). We then evaluate the residues in \eqref{eq:AI line 5}. In \eqref{eq:AI line 6}, we use the global $J_0^3$ Ward identity \eqref{eq:global Ward identities} on the first term. Using \eqref{eq:Fiwi}, the second term also equals an affine primary correlation function. Thus, both terms do not contain any derivatives of $\delta$ functions and hence vanish when multiplied with the constraint $f_I$. 

We thus conclude $A_I=0$ and so
\be 
G_I(z)=f_I G(z)=0\ .
\ee
Hence $G(z)$ vanishes when being multiplied by any of the contraints $f_I$. This implies that $G(z)$ does not contain any derivatives of $\delta$ functions. We should also note that this proof also works backwards: requiring that $G_I(z)=0$ is equivalent to demanding the constraint equations \eqref{eq:constraint equations} for the derivative pieces.
\subsubsection{Step 2: Solving the \texorpdfstring{$\delta'$}{delta'} pieces}\label{subsubsec:step 2}
We can now solve the constraint equations completely for the terms that contain a derivative $\delta'$. We claim that
\begin{align}
f_I\left\langle J^+(z) \prod_{i=1}^n V_{h_i}^{w_i}(x_i,z_i) \right\rangle=\frac{f_I}{\partial \Gamma(z)} Q(z)\ , \label{eq:Jp deltap solution}
\end{align}
where $Q(z)$ is a quadratic differential on the punctured Riemann surface $\Sigma_{g,n}$, i.e.~a quadratic differential on $\Sigma_g$ that has at most a single pole at the insertion points $z_i$. We furthermore require that the quadratic differential satisfies Ward identities, i.e.
\begin{align}
\sum_{i=1}^n x_i^{m+1} \partial_{x_i} Q(z)=0
\end{align}
for $m=-1,0,1$. The space of quadratic differentials on the punctured Riemann surface has dimension $3g-3+n$, since it parametrises the cotangent space to the surface in the moduli space $\mathcal{M}_{g,n}$. These $3g-3+n$ parameters are tuned to satisfy the input conditions of the constraint equations, namely the residues and the zero mode insertions of the currents.

We should note that this solution has a pole of order $w_i$ at the insertion points $z_i$. This is the correct order, since while the OPE of $J^+(z)$ with the affine primary field $V_{h_i}^{w_i}(x_i,z_i)$ has a pole of order $w_i+1$, the highest order pole is $F_{w_i}^i$, which according to \eqref{eq:Fiwi} can be written as a correlation function of affine primaries and hence does not contain derivatives of $\delta$ functions.

Using the transformation properties of the covering map under M\"obius transformations \eqref{eq:constraints Moebius invariance} and the Ward identity \eqref{eq:relation between current insertions}, one can derive the form of the correlators with $J^3(z)$ and $J^-(z)$ insertions.\footnote{Note that applying \eqref{eq:relation between current insertions} to terms with no $\delta$ derivative only yields terms with no $\delta$ derivative and thus one can apply it on the derivative terms separately.} They take the form
\begin{subequations}\label{eq:J3 and Jm deltap solutions}
\begin{align}
f_I\left\langle J^3(z) \prod_{i=1}^n V_{h_i}^{w_i}(x_i,z_i) \right\rangle&=\frac{f_I \Gamma(z)}{\partial \Gamma(z)}Q(z)\ , \\
f_I\left\langle J^-(z) \prod_{i=1}^n V_{h_i}^{w_i}(x_i,z_i) \right\rangle&=\frac{f_I \Gamma(z)^2}{\partial \Gamma(z)} Q(z)\ .
\end{align}
\end{subequations}
It hence follows immediately that
\be 
f_I G(z)=0\ ,
\ee
which ensures that all the constraint equations for the derivative terms are satisfied. Thus, we have shown that this is the correct solution that follows from the constraint equations.
\subsubsection{Step 3: \texorpdfstring{$H(z)$}{H(z)} has no \texorpdfstring{$\delta'$}{delta'} piece}\label{subsubsec:step 3}
Let us now consider $H(z)$ as defined in eq.~\eqref{eq:H definition}.
From the solution of Step~\ref{subsubsec:step 2}, it follows immediately that $f_I H(z)=0$. Hence also $H(z)$ contains no derivatives of $\delta$ functions.
\subsubsection{Step 4: \texorpdfstring{$G(z)$}{G(z)} is independent of the choice of \texorpdfstring{$\Gamma(z)$}{Gamma(z)}}\label{subsubsec:step 4}
Next, we show that our definition of $G(z)$ \eqref{eq:G definition} is independent of the choice of the choice of extension of $\Gamma(z)$. For this, let $\Gamma^{(1)}(z)$ and $\Gamma^{(2)}(z)$ be two possible extensions of the covering map and let $G^{(1)}(z)$ and $G^{(2)}(z)$ be the corresponding $G(z)$'s. We then compute
\begin{align}
G^{(1)}(z)-G^{(2)}(z)&=-2(\Gamma^{(1)}(z)-\Gamma^{(2)}(z))H(z)\ .
\end{align}
The prefactor $(\Gamma^{(1)}(z)-\Gamma^{(2)}(z))$ vanishes on the localisation locus and hence the right hand side is a linear combination of terms of the form $f_I H(z)$. But since $f_I H(z)=0$, the right hand side vanishes. Thus,
\be 
G^{(1)}(z)=G^{(2)}(z)\ .
\ee
This shows that $G(z)$ is independent of the choice of $\Gamma(z)$.
\subsubsection{Step 5: \texorpdfstring{$G(z)$}{G(z)} is proportional to \texorpdfstring{$\partial \Gamma(z)$}{dGamma(z)}}\label{subsubsec:step 5}
We inspect the Taylor expansion of $G(z)$ around $z=z_m$. Since $G(z)$ is independent of the choice of $\Gamma(z)$, we can choose $\Gamma(z)$ conveniently by requiring that the Taylor expansion \eqref{eq:Gamma Taylor expansion} also holds away from the localisation locus for $z_m$ (but not for the other $z_j$'s).
We then obtain for $r=0,\dots,w_m-1$
\begin{align}
\frac{\partial_z^r}{r!} G(z=z_m)&=\frac{\partial_z^r}{r!}\left\langle\left( J^-(z)-2\Gamma(z) J^3(z)+\Gamma(z)^2 J^+(z)\right) \prod_{j=1}^n V_{h_j}^{w_j}(x_j,z_j) \right \rangle\nonumber\\
&\qquad+\delta_{r,w_m-1}\left(a_m^2  F_{w_m}^m-2 a_m h_m \left\langle \prod_{j=1}^n V_{h_j}^{w_j}(x_j,z_j) \right \rangle\right)\\
&\overset{!}{=}\delta_{r,w_m-1}\Bigg(\left(h_m-\tfrac{k+2}{2}w_m-j_m\right)\left \langle  V_{h_m-1}^{w_m}(x_m,z_m)   \prod_{j\ne m}^n V_{h_j}^{w_j}(x_i,z_j) \right \rangle\nonumber\\
&\qquad+a_m^2  F_{w_m}^m-2 a_m h_m \left\langle \prod_{j=1}^n V_{h_j}^{w_j}(x_j,z_j) \right \rangle\Bigg)\ . \label{eq:G expansion}
\end{align}
Thus, the derivatives vanish for $r=0,\dots,w_m-2$, while asking equality of the $(w_m-1)$-th derivative with the right hand side is the recursion relation that we will analyse below.

From the definition of $G(z)$ \eqref{eq:G definition}, it is also clear that $G(z)$ has double poles located at the poles of the covering map, but has no other poles.

Thus, we obtain that
\be 
\frac{G(z)}{\partial \Gamma(z)}
\ee 
is a function on the Riemann surface that has no poles in $z$ and is hence independent of $z$. We thus have
\be 
G(z)=A \partial \Gamma(z)
\ee
for some constant $A$ (that of course still depends on all the other parameters in the problem).
\subsubsection{Step 6: Computing the proportionality factor}\label{subsubsec:step 6}
We want to compute the proportionality factor $A$. For this, we follow essentially the same steps as for the computation of $A_I$ above. One has to be however more careful, since derivatives of $\delta$ functions can be present. For this computation, it is convenient to make a definite choice for the extension of the covering map, which we define as follows. We only require that
\be 
\Gamma(z_i)=x_i
\ee 
for $i=1,2,3$, but not for $i \ge 4$.\footnote{Requiring $\Gamma(z_i)=x_i$ for three points is always possible by the M\"obius covariance discussed in Section~\ref{subsubsec:SL2R covariance}.} We moreover allow the covering map to be arbitrarily twisted by group elements $\gamma_1,\dots,\gamma_g$ along the $\beta_\mu$ cycles., i.e.
\be 
\Gamma(\beta_\mu(z))=\gamma_\mu(\Gamma(z))\ .
\ee
This introduces $3g-3+n$ free parameters in the definition and by dimension counting, one expects that such an extension always exists, at least in the vicinity of the localisation locus. We then have
\begin{align}
A N &=-\sum_{a=1}^N \mathop{\text{Res}}_{z=z_a^*} \frac{G(z)}{\Gamma(z)} \label{eq:A line 1}\\
&=-\sum_{a=1}^N \mathop{\text{Res}}_{z=z_a^*}  \Gamma(z)\left\langle J^+(z) \prod_{j=1}^n V_{h_j}^{w_j}(x_j,z_j) \right \rangle\label{eq:A line 2}\\
&=\sum_{i=1}^n\mathop{\text{Res}}_{z=z_i}  \Gamma(z)\left\langle J^+(z) \prod_{j=1}^n V_{h_j}^{w_j}(x_j,z_j) \right \rangle\nonumber\\
&\qquad-\frac{1}{2\pi i} \oint_\mathcal{C} \Gamma(z)\left\langle J^+(z) \prod_{j=1}^n V_{h_j}^{w_j}(x_j,z_j) \right \rangle\label{eq:A line 3}\ ,
\end{align}
where $\mathcal{C}$ is the contour \eqref{eq:cut contour} that bounds the fundamental domain of the Riemann surface. Since our choice of $\Gamma(z)$ was not periodic, this term appears additionally. Let us first evaluate the contour integral further
\begin{align}
-\frac{1}{2\pi i} \oint_\mathcal{C}  &\ \Gamma(z)\left\langle J^+(z) \prod_{j=1}^n V_{h_j}^{w_j}(x_j,z_j) \right \rangle\nonumber\\
&=\frac{1}{2\pi i}\sum_{\mu=1}^g \oint_{\alpha_\mu} (\Gamma(\beta_\mu(z))-\Gamma(z))\left\langle J^+(z) \prod_{j=1}^n V_{h_j}^{w_j}(x_j,z_j) \right \rangle\nonumber\\
&=\frac{1}{2\pi i}\sum_{\mu=1}^g \oint_{\alpha_\mu} (\gamma_\mu(\Gamma(z))-\Gamma(z))\left\langle J^+(z) \prod_{j=1}^n V_{h_j}^{w_j}(x_j,z_j) \right \rangle\ .
\end{align}
We now use that the correlator is localising and thus we only need to expand $\gamma_\mu$ up to first order, i.e. 
\be 
\gamma_\mu(x)=x+\varepsilon^+_\mu +\varepsilon^3_\mu x+\varepsilon^-_\mu x^2\ .
\ee
Thus,
\begin{align}
-\frac{1}{2\pi i} \oint_\mathcal{C}  &\ \Gamma(z)\left\langle J^+(z) \prod_{j=1}^n V_{h_j}^{w_j}(x_j,z_j) \right \rangle\nonumber\\
&=\frac{1}{2\pi i}\sum_{\mu=1}^g \oint_{\alpha_\mu} (\varepsilon^+_\mu+\varepsilon^3_\mu \Gamma(z)+\varepsilon^-_\mu \Gamma(z)^2)\left\langle J^+(z) \prod_{j=1}^n V_{h_j}^{w_j}(x_j,z_j) \right \rangle\label{eq:simplifying contour 1}\\
&=\frac{1}{2\pi i}\sum_{\mu=1}^g \oint_{\alpha_\mu} \Bigg(\varepsilon^+_\mu\left\langle J^+(z) \prod_{j=1}^n V_{h_j}^{w_j}(x_j,z_j) \right \rangle+\varepsilon^3_\mu\left\langle J^3(z) \prod_{j=1}^n V_{h_j}^{w_j}(x_j,z_j) \right \rangle\nonumber\\
&\qquad+\varepsilon^-_\mu\left\langle J^-(z) \prod_{j=1}^n V_{h_j}^{w_j}(x_j,z_j) \right \rangle\Bigg)\label{eq:simplifying contour 2}\\
&=\frac{1}{2\pi i}\sum_{\mu=1}^g\Bigg(\varepsilon^+_\mu\left\langle J^+_{0,\mu} \prod_{j=1}^n V_{h_j}^{w_j}(x_j,z_j) \right \rangle+\varepsilon^3_\mu\left\langle J^3_{0,\mu} \prod_{j=1}^n V_{h_j}^{w_j}(x_j,z_j) \right \rangle\nonumber\\
&\qquad+\varepsilon^-_\mu\left\langle J^-_{0,\mu} \prod_{j=1}^n V_{h_j}^{w_j}(x_j,z_j) \right \rangle\Bigg)\ .\label{eq:simplifying contour 3}
\end{align}
Here, we used that $\varepsilon_\mu^a$ vanishes when the localisation constraint is satisfied and hence we can use the solution \eqref{eq:Jp deltap solution} and \eqref{eq:J3 and Jm deltap solutions} for the $\delta'$ pieces to simplify the result.
Finally, we discuss the three terms. For this, we note that we can choose the $3g-3+n$ constraints $f_I$ as 
\be 
\{\varepsilon_\mu^a\}_{\mu=1,\dots,g,\, a=+,-,3} \qquad \text{and} \qquad \{\Gamma(z_i)-x_i\}_{i=4,\dots,n}\ . 
\ee
We have then by definition \eqref{eq:definition zero mode insertion}
\be 
\left \langle J^a_{\mu,0} \prod_{j=1}^n V_{h_j}^{w_j}(x_j,z_j) \right \rangle= -2\pi i \partial_{\varepsilon_\mu^a} \left \langle \prod_{j=1}^n V_{h_j}^{w_j}(x_j,z_j) \right \rangle\Bigg|_{\varepsilon^a_\mu=0}\ .
\ee
Hence, using \eqref{eq:delta distributional identity}, it follows that
\be 
-\frac{1}{2\pi i} \oint_\mathcal{C}  \ \Gamma(z)\left\langle J^+(z) \prod_{j=1}^n V_{h_j}^{w_j}(x_j,z_j) \right \rangle=3g \left\langle  \prod_{j=1}^n V_{h_j}^{w_j}(x_j,z_j) \right \rangle\ ,
\ee
since every of the term in the sum of  \eqref{eq:simplifying contour 3} contributes $-2\pi i$.

Continuing the computation of \eqref{eq:A line 3}, we have
\begin{align}
AN&=\sum_{i=1}^n\mathop{\text{Res}}_{z=z_i}  \ (\Gamma(z_i)+a_i^\Gamma(z-z_i)^{w_i})\left\langle J^+(z) \prod_{j=1}^n V_{h_j}^{w_j}(x_j,z_j) \right \rangle\nonumber\\
&\qquad+3g\left\langle \prod_{j=1}^n V_{h_j}^{w_j}(x_j,z_j) \right \rangle\label{eq:A line 4}\\
&=\sum_{i=1}^n (\Gamma(z_i) F_0^i+a_i^\Gamma F_{w_i}^i)+3g\left\langle \prod_{j=1}^n V_{h_j}^{w_j}(x_j,z_j) \right \rangle\label{eq:A line 5}\\
&=\sum_{i=1}^n (x_i F_0^i+a_i^\Gamma F_{w_i}^i)+3g\left\langle \prod_{j=1}^n V_{h_j}^{w_j}(x_j,z_j) \right \rangle\nonumber\\
&\qquad+\sum_{i=4}^n (\Gamma(z_i)-x_i)\partial_{x_i}\left\langle \prod_{j=1}^n V_{h_j}^{w_j}(x_j,z_j) \right \rangle\label{eq:A line 6} \\
&=\sum_{i=1}^n a_i^\Gamma F_{w_i}^i+\left(3g+n-3-\sum_{i=1}^n h_i \right)\left\langle \prod_{j=1}^n V_{h_j}^{w_j}(x_j,z_j) \right \rangle\ .\label{eq:A line 7}
\end{align}
Here, we followed the same steps as in the derivation of $A_I$ above. There is an extra term appearing, since $\Gamma(z_i)$ for $i \ge 4$ is not equal to $x_i$. See also \cite{Eberhardt:2019ywk}, where this was discussed for the sphere. We should also mention that we could insert for $a_i^\Gamma$ its value for the true covering map, since it multiplies $F_{w_i}^i$, which does not possess derivatives of delta functions.
\subsubsection{Step 7: The recursion relations}\label{subsubsec:step 7}
Finally, we derive the recursion relations under the assumption that the solution is localising. This is done by expanding $G(z)$ around $z_m$ and extracting the coefficient of the term $(z-z_m)^{w_m}$. We compute
\be 
G(z)=A \partial \Gamma(z)=A\, a_m^\Gamma w_m (z-z_m)^{w_m-1}+\mathcal{O}((z-z_m)^{w_m})\ .
\ee
Imposing equality with the right-hand side of \eqref{eq:G expansion} yields the recursion relations, which take the form
\begin{multline}
(a_m^\Gamma)^{-1} \left(h_m-\tfrac{k+2}{2}w_m-j_m \right) \left\langle V_{h_m-1}^{w_m}(x_m,z_m) \prod_{j\ne m}^n V_{h_j}^{w_j}(x_j,z_j) \right \rangle\\
=\sum_{i=1}^n \left(\frac{w_m}{N}- \delta_{i,m}\right)a_i^\Gamma\left(h_i-\tfrac{k+2}{2}w_i+j_i \right) \left\langle V_{h_i+1}^{w_i}(x_i,z_i) \prod_{j\ne i}^n V_{h_j}^{w_j}(x_j,z_j) \right \rangle\\
+\left(\frac{w_m}{N}\left(3g-3+n-\sum_{i=1}^n h_i \right)+2h_m\right)  \left\langle \prod_{j=1}^n V_{h_j}^{w_j}(x_j,z_j) \right \rangle\ , \label{eq:localised recursion relations}
\end{multline}
where $N$ is given by the Riemann Hurwitz formula \eqref{eq:Riemann Hurwitz}.

Finally, we have to check that the localising solution \eqref{eq:localising solution} indeed solves these recursion relations. This is straightforward to check and one finds that it is the case, provided that the constraint \eqref{eq:j constraint} on the spins is satisfied.

This concludes the proof that the localising solution \eqref{eq:localising solution} indeed solves the recursion relations.
\section{Discussion} \label{sec:discussion}
We have studied correlation functions of the $\mathrm{SL}(2,\mathds{R})_{k+2}$ WZW model on higher genus Riemann surfaces. We have in particular included spectrally flown vertex operators in our analysis. Let us summarise our most important findings.
\smallskip

We have derived the local Ward identities \eqref{eq:local Ward identities} for the correlators that relate correlators with current insertions to correlators without current insertions, but zero-mode insertions that introduce twisted boundary conditions along the cycles of the Riemann surface. To fully solve the local Ward identities in the spectrally flown case, they have to be complemented by the constraint equations \eqref{eq:constraint equations} that allow one to solve for all the unknowns that appear on the right-hand-side of the Ward identities.

We subsequently used these local Ward identities to derive a constraint on the correlators of spectrally flown affine primary fields. This constraint takes the form of recursion relations relating correlators with amounts of $\mathrm{SL}(2,\mathds{R})$ conformal weight, see \eqref{eq:recursion relation schematic form} for a schematic version of the recursion relations. We have not been able to derive them in full generality. We pointed out that these recursion relations always possess a very simple natural solution that localises on a finite subset in the moduli space of Riemann surfaces corresponding to the covering surface of a given $n$-punctured Riemann sphere on which the dual CFT is defined. Given the complexity of the recursion relations, the existence of this simple solution makes a strong case for its correctness.

Given the localising solution, we have explained in Section~\ref{sec:exact AdS3CFT2 correspondence} that this makes the equivalence of the worldsheet theory with the dual CFT almost manifest and relates the genus expansion on both sides of the duality directly. We are hence confident that a perturbative proof of the $\mathrm{AdS}_3/\mathrm{CFT}_2$ correspondence in the instance of the tensionless limit of $\mathrm{AdS}_3 \times \mathrm{S}^3 \times \mathbb{T}^4$ is possible and we have taken another step in this direction.
\medskip

Let us now discuss various aspects of our construction and future directions.

\paragraph{Unique solution.} We have shown that the localising solution \eqref{eq:localising solution} is \emph{a} solution to the recursion relations, but we have not shown that it is the unique solution. This is presumably not the case and to show uniqueness of the solution one would need a further input that constrains correlators. There are two sources of such additional constraints. 

For $k=1$, one can study the worldsheet theory directly in the hybrid formalism, as was done in \cite{Eberhardt:2019ywk}. The WZW model $\mathrm{PSU}(1,1|2)_1$ has additional null-vectors in its vacuum module and this is what restricted the representations to all have $\mathrm{SL}(2,\mathds{R})$ spin $\tfrac{1}{2}$. Hence one expects that these additional null-vectors restrict the correlators further. This is explored in \cite{AndreaMatthiasRajesh}.

For general $k$, correlators are further restricted by the mixed Ward-identities, i.e.~the Knizhnik-Zamolodchikov equations. They relate world-sheet coordinate dependence with spacetime coordinate dependence. One could hope that all these constraints together can determine the correlators completely (or rather up to an overall constant, which would need to be fixed by crossing symmetry). 

\paragraph{Why does the $\mathrm{SL}(2,\mathds{R})$ WZW model encode covering maps?} It is very striking from a physical point of view that the $\mathrm{SL}(2,\mathds{R})$ WZW model seems to have information about covering maps encoded in its correlation functions. While the proof of Section~\ref{subsec:proof} explains in principle where this comes from, one would like to make the connection more manifest. The theory has many aspects of a topological A-model, except that it counts holomorphic maps to the boundary of $\mathrm{AdS}_3$ and not to $\mathrm{AdS}_3$ itself.

\paragraph{Correlators away from $j$ constraint.} While for the tensionless limit $k=1$, the constraint \eqref{eq:j constraint} on $j$ is always satisfied, this is not true for the $\mathrm{SL}(2,\mathds{R})$ WZW model in general. One would like to understand the solutions of the recursion relations away from the constraint on $j$, which brings one closer towards solving the $\mathrm{SL}(2,\mathds{R})$ WZW model.

\paragraph{Schottky parametrisation.} The result we have derived is conceptually clear, but relatively abstract. One could try to make it more concrete by choosing an explicit realisation of the moduli space of Riemann surfaces of genus $g$. A parametrisation that is especially suited for exposing the structure of moduli space is the Schottky parametrisation, where one constructs genus $g$ surfaces as a quotient of the complex plane by a freely acting subgroup of $\mathrm{SL}(2,\mathds{C})$. This group is generated by $g$ group elements of $\mathrm{SL}(2,\mathds{C})$ subject to certain conditions and up to overall conjugation, which leads to a $3g-3$-dimensional moduli space. This space was used efficiently in \cite{Bernard:1988yv} to construct the relevant functions explicitly on higher genus surfaces.

\paragraph{Boundary higher genus.} We have in this paper considered the theory on a higher-genus worldsheet. One could also try to put the boundary of $\mathrm{AdS}_3$ on a higher-genus surface. The case of a boundary torus is thermal $\mathrm{AdS}_3$ and was considered in the literature before \cite{Maldacena:2000kv, Eberhardt:2018ouy}, although we are not aware of a construction that goes beyond the thermal partition function. It is well-known that the thermal partition function of the $\mathrm{SL}(2,\mathds{R})$ WZW-model has poles in the moduli space of tori that correspond to covering surfaces of the boundary torus \cite{Maldacena:2000kv}. This statement becomes sharper for the $k=1$ theory of strings on $\mathrm{AdS}_3 \times \mathrm{S}^3 \times \mathbb{T}^4$, where the thermal partition function is $\delta$-function localised on holomorphic maps \cite{Eberhardt:2018ouy}. We hence expect that one can define a version of the $\mathrm{SL}(2,\mathds{R})$ WZW model with higher-genus boundary. For the tensionless limit on $\mathrm{AdS}_3 \times \mathrm{S}^3 \times \mathbb{T}^4$, we expect that correlators would again localise on covering surfaces of the boundary higher-genus surface. It would be very interesting to give such a definition and explore it. 

\paragraph{Topological version and Hurwitz theory.} While we have argued that the worldsheet theory possesses many properties that resemble a topological string theory, it is still a physical string model. To sidestep the difficulties of the physical string one could try to consider a topological version of the theory, for instance by topologically twisting string theory on $\mathrm{AdS}_3 \times \mathrm{S}^3$ \cite{Rastelli:2005ph}, see also \cite{Costello:2020jbh} for a recent twisted version of the correspondence in the supergravity limit. 

\paragraph{Higher-dimensional versions and Gromov-Witten theory.} One might wonder whether it is in general possible to construct other 2d CFTs with similar properties. For instance, does the WZW model based on $\mathrm{SU}(1,2)$ carry similar information about the holomorphic maps from the worldsheet to the boundary of the group, the Hermitian symmetric space $\mathrm{SU}(1,2)/(\mathrm{S}(\mathrm{U}(1) \times \mathrm{U}(2)))$? Hurwitz theory is equivalent to Gromov-Witten theory with curves as target spaces \cite{Okounkov:2006}, so it is natural to suspect that possible generalisations count holomorphic maps to the boundary of the group and are hence related to Gromov-Witten theory. 
\section*{Acknowledgements}
I would like to thank Andrea Dei, Matthias Gaberdiel, Rajesh Gopakumar, Zohar Komargodski, Shota Komatsu, Sebastian Mizera, Tom\'{a}\v{s} Proch\'{a}zka, Mukund Rangamani, Leonardo Rastelli and Edward Witten for enlightening discussions. 
I also thank Andrea Dei, Matthias Gaberdiel and Rajesh Gopakumar for helpful comments on the manuscript. 
I acknowledge the hospitality of the Simons Center for Geometry and Physics towards the end of this project. I thank the della Pietra family for support at the IAS. 
\appendix
\section{Some background on Riemann surfaces} \label{app:Riemann surfaces background}
In this appendix, we review some useful concepts from the theory of (compact) Riemann surfaces. We explain in particular the construction of the kernel $\Delta(z,\zeta)$.
\subsection{The period matrix}\label{subapp:period matrix}
Recall that a Riemann surface of genus $g$ has $g$ independent holomorphic differentials. We can moreover choose a canonical basis $\alpha_1,\dots,\alpha_g,\beta_1,\dots,\beta_g$ for the homology $\mathrm{H}_1(\Sigma_{g},\mathds{Z})$, in which the intersection product takes the form
\be 
\alpha_\mu \cap \alpha_\nu=0\ , \qquad \beta_\mu \cap \beta_\nu=0\ , \qquad \alpha_\mu \cap \beta_\nu=\delta_{\mu\nu}\ .
\ee
Such a basis is unique up to $\mathrm{Sp}(2g,\mathds{Z})$ transformations. A Riemann surface together with such a choice of homology basis is called marked Riemann surface and in the main text, we fix one particular marking.

\begin{figure}
\begin{center}
\begin{tikzpicture}
\draw[smooth, thick] (0,1) to[out=30,in=150] (2,1) to[out=-30,in=210] (3,1) to[out=30,in=150] (5,1) to[out=-30,in=30] (5,-1) to[out=210,in=-30] (3,-1) to[out=150,in=30] (2,-1) to[out=210,in=-30] (0,-1) to[out=150,in=-150] (0,1);
\draw[smooth, thick] (0.4,0.1) .. controls (0.8,-0.25) and (1.2,-0.25) .. (1.6,0.1);
\draw[smooth, thick] (0.5,0) .. controls (0.8,0.2) and (1.2,0.2) .. (1.5,0);
\draw[smooth, thick] (3.4,0.1) .. controls (3.8,-0.25) and (4.2,-0.25) .. (4.6,0.1);
\draw[smooth, thick] (3.5,0) .. controls (3.8,0.2) and (4.2,0.2) .. (4.5,0);
\draw[thick,red] (2.5,0) .. controls (-1,2) and (-1,-2) .. (2.5,0);
\draw[thick,red] (2.5,0) .. controls (6,2) and (6,-2) .. (2.5,0);
\draw[thick,blue,bend right=20] (2.5,0) to (1.5,0);
\draw[thick,blue, dashed,bend right=50] (1.5,0) to (2,-1);
\draw[thick,blue,bend right=40] (2,-1) to (2.5,0);
\draw[thick,blue,bend left=20] (3.5,0) to (2.5,0);
\draw[thick,blue, dashed,bend right=50] (3.5,0) to (3,1);
\draw[thick,blue,bend right=40] (3,1) to (2.5,0);
\node at (1,.8) {$\alpha_1$};
\node at (4,-.8) {$\alpha_2$};
\node at (2.3,.5) {$\beta_1$};
\node at (2.7,-.5) {$\beta_2$};
\end{tikzpicture}
\end{center}
\caption{The choice of homology basis of the Riemann surface for the example $g=2$.}\label{fig:homology basis}
\end{figure}
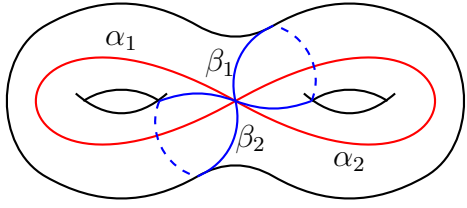

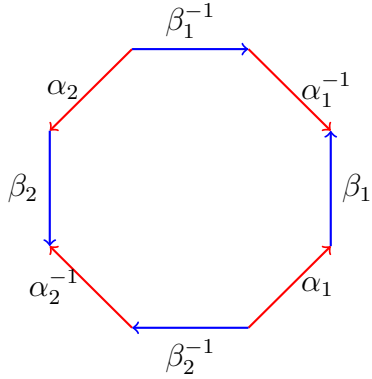
\begin{figure}
\begin{center}
\begin{tikzpicture}
\draw[thick,<-,red] (22.5:2) to node[right,black] {$\alpha_1^{-1}$} (67.5:2);
\draw[thick,->,red] (112.5:2) to node[left,black] {$\alpha_2$} (157.5:2);
\draw[thick,<-,red] (202.5:2)  to node[left,black] {$\alpha_2^{-1}$}  (247.5:2); 
\draw[thick,->,red] (292.5:2) to node[right,black] {$\alpha_1$} (337.5:2);
\draw[thick,<-,blue] (67.5:2) to node[above,black] {$\beta_1^{-1}$} (112.5:2);
\draw[thick,->,blue] (157.5:2) to node[left,black] {$\beta_2$} (202.5:2);
\draw[thick,<-,blue] (247.5:2)  to node[below,black] {$\beta_2^{-1}$}  (292.5:2); 
\draw[thick,->,blue] (337.5:2) to node[right,black] {$\beta_1$} (22.5:2);
\end{tikzpicture}
\end{center}
\caption{The construction of a Riemann surface by gluing, here drawn for the example $g=2$.} \label{fig:cut surface}
\end{figure}

We can moreover choose the homology basis such that all $\alpha_\mu$'s intersect transversely with $\beta_\nu$ in one single point $z_0$. This is illustrated in Figure~\ref{fig:homology basis}.
We can in particular realise the Riemann surface topologically by a quotient of its universal covering space $\widetilde{\Sigma}_g$ in which the sides of a $4g$-gon are glued as illustrated in Figure~\ref{fig:cut surface}. This follows by construction from the choice of cycles as in Figure~\ref{fig:homology basis}.

We fix a canonical basis $\omega_1,\dots,\omega_g$ of the holomorphic differentials by imposing that
\be 
\int_{\alpha_\mu}\omega_\nu=\delta_{\mu\nu}\ .
\ee
The period matrix of the Riemann surface is the $g \times g$ matrix defined by
\be 
\boldsymbol{\Omega}_{\mu\nu}=\int_{\beta_\mu}\omega_\nu\ .
\ee
The Riemann bilinear relations state that 
\be 
\boldsymbol{\Omega}\tran=\boldsymbol{\Omega}\quad \text{and}\quad \text{Im}(\boldsymbol{\Omega})>0\ . \label{eq:Riemann bilinear relations}
\ee
\subsection{The Jacobian and the Abel map}\label{subapp:jacobian and abel}
The Jacobian associated to the Riemann surface $\Sigma_g$ is the complex $g$-dimensional torus $\mathbb{T}^{2g}$ given by
\be 
 \mathrm{Jac}=\mathds{C}^g/(\mathds{Z}^g\oplus \boldsymbol{\Omega}\, \mathds{Z}^g)\ .
\ee
The Abel map is the canonical embedding of the Riemann surface into its Jacobian given by
\begin{subequations}
\begin{align}
\mathrm{Ab}:\ \Sigma_g &\longrightarrow \mathrm{Jac}\ , \\
z&\longmapsto \mathbf{z}=\Bigg(\int_{z_0}^z \omega_1\, ,\ \dots\, , \ \int_{z_0}^z \omega_g\Bigg)\ .
\end{align}
\end{subequations}
Here, $\int_{z_0}^z$ is the integral along an arbitrary path connecting $z_0$ and $z$, where $z_0$ is some fixed point on the Riemann surface. Clearly, the definition is independent of the choice of the path.
We will in the following denote by $\mathbf{z}$ the image of the point $z$ under the Abel map. The Abel map can also be lifted to the universal covering spaces $\widetilde{\Sigma}_g$ and $\mathds{C}^g$ and we continue to denote the image of the point $z$ by $\mathbf{z}$.

The Abel map is injective for $g \ge 1$ and an isomorphism for $g=1$.
\subsection{The theta series}\label{subapp:theta}
We define the theta series $\vartheta_{\alpha,\beta}(\mathbf{z}\, |\, \boldsymbol{\Omega})$ as
\be 
\vartheta_{\alpha,\beta}(\mathbf{z}\, |\, \boldsymbol{\Omega})\equiv\sum_{\mathbf{n}\in \mathds{Z}^g+\alpha} \mathrm{e}^{\pi i \mathbf{n}\tran \boldsymbol{\Omega}\mathbf{n}+2\pi i \mathbf{n}\tran (\mathbf{z}+\beta)}\ ,
\ee
where $\alpha$, $\beta \in \mathds{R}^g/\mathds{Z}^g$ and $\mathbf{z}\in \mathds{C}^g$. This sum is absolutely converging, because of the Riemann bilinear relations \eqref{eq:Riemann bilinear relations}.

We choose in the following $\alpha$ and $\beta$ such that
\be 
2\alpha=2\beta=0\quad\text{and}\quad \alpha\tran \beta\in \tfrac{1}{4}+\tfrac{1}{2}\mathds{Z}
\ee
and denote $\vartheta_{\alpha,\beta}(\mathbf{z}\, |\, \boldsymbol{\Omega})$ resulting from such a choice just by $\vartheta(\mathbf{z}\, |\, \boldsymbol{\Omega})$. Such a choice is always possible. We note that with this choice, $\vartheta(\mathbf{0}\, |\, \boldsymbol{\Omega})=0$, since the term $\boldsymbol{n}$ in the sum cancels with $-\mathbf{n}$.

$\vartheta(\mathbf{z}\, |\, \boldsymbol{\Omega})$ has the following quasi-periodicity property: for $\mathbf{m} ,\, \mathbf{n}\in \mathds{Z}^g$, we have
\be 
\vartheta(\mathbf{z}+\mathbf{m}+\boldsymbol{\Omega}\, \mathbf{n}\, |\, \boldsymbol{\Omega})=\mathrm{e}^{-\pi i \mathbf{n}\tran \boldsymbol{\Omega}\mathbf{n}-2\pi i \mathbf{n}\tran(\mathbf{z}+\beta)+2\pi i \mathbf{m}\tran\alpha} \vartheta(\mathbf{z}\, |\, \boldsymbol{\Omega})\ . \label{eq:theta series quasi periodicity}
\ee
We denote similarly 
\be 
\vartheta(z_1,\, z_2\, | \, \boldsymbol{\Omega})\equiv \vartheta(\mathbf{z}_2- \mathbf{z}_1\, | \, \boldsymbol{\Omega})=\vartheta\Big(\int_{z_1}^{z_2} \omega_1,\, \dots,\, \int_{z_1}^{z_2} \omega_g\, \Big| \, \boldsymbol{\Omega}\Big)
\ee
for two points $z_1$ and $z_2$ in the universal covering space $\widetilde{\Sigma}_g$.

$\vartheta(z_1,\, z_2\, |\, \boldsymbol{\Omega})$ satisfies the analogue of quasi-periodicity properties \eqref{eq:theta series quasi periodicity} around the $2g$ cycles of the surface. Moreover $\vartheta(z,\, z\, |\, \boldsymbol{\Omega})=0$. 

$\vartheta(z_1,\, z_2\, |\, \boldsymbol{\Omega})$ is known as the prime form on the Riemann surface in the mathematical literature. Its significance is that it serves as a basis to construct all meromorphic functions on a Riemann surface.
An important property of $\vartheta(z_1,\, z_2\, |\, \boldsymbol{\Omega})$ are its zeros. Even though $\vartheta(z_1,\, z_2\, |\, \boldsymbol{\Omega})$ is only quasi-periodic, its zeros are periodic and hence well-defined in $\Sigma_g$, since the prefactor in \eqref{eq:theta series quasi periodicity} is non-zero. The zeros for $z_1,\,z_2 \in \Sigma_g$ are
\be 
\vartheta(z_1,\, z_2\, |\, \boldsymbol{\Omega})=0 \quad \Longleftrightarrow z_1=z_2\text{ or } z_1=P_1,\dots,P_{g-1}\text{ or } z_2=Q_1,\dots,Q_{g-1}\ .
\ee
Here, the points $P_i$ ($Q_i$) are independent of $z_2$ ($z_1$).\footnote{Sometimes $\vartheta(z_1,\, z_2\, |\, \boldsymbol{\Omega})$ is referred to as the prime function. The prime form then refers to a bi-half-differential, where the additional zeros $P_i$ and $Q_i$ are cancelled. We will not have need for this quantity. See e.g.~\cite{Fay:1973}.}
\subsection{The logarithmic derivative of the prime form}\label{subapp:logarithmic derivative prime form}
In the main text, a very important role is played by the logarithmic derivative of the prime form,
\be 
\Delta(z,\zeta\, |\, \boldsymbol{\Omega})\equiv \partial_{z} \, \log\big(\vartheta(z,\zeta\, |\, \boldsymbol{\Omega})\big)\ .
\ee
It is a function in $\zeta$ and a one-form in $z$ on the universal covering space $\widetilde{\Sigma}_g$. The construction of $\Delta$ depends on a marking, i.e.~a choice of $\alpha_\mu$ and $\beta_\mu$ and a choice of the characteristics $\alpha$ and $\beta$ in the theta-function.
Its periodicity properties are given in eq.~\eqref{eq:Delta quasi periodicity}, they follow directly from \eqref{eq:theta series quasi periodicity}.

$\Delta(\zeta\, ,z\, |\, \boldsymbol{\Omega})$ has the following poles as a result of the singularities of the zeros of $\vartheta(z,\zeta\, |\, \boldsymbol{\Omega})$,
\be 
z=\zeta,\, Q_1,\,\dots,\, Q_{g-1}\ ,
\ee
all with residue one. We stress again that $Q_1,\, \dots,\, Q_{g-1}$ are independent of $\zeta$. The only pole in $\zeta$ occurs at $\zeta=z$.

\section{Elliptic functions} \label{app:elliptic functions}
In this appendix, we review some facts about elliptic functions, that are needed to check the examples of covering maps of Section~\ref{subsubsec:examples}.
\subsection{The Eisenstein Series} \label{subapp:Eisenstein series}
We define Eisenstein series for $k \in \mathds{N}$ as
\begin{align}
G_{2k}(\tau)&=2\zeta(2k) E_{2k}(\tau)\ ,\\
E_{2k}(\tau)&=1-\frac{4k}{B_{2k}}\sum_{n=1}^\infty \sigma_{2k-1}(n)q^n\ ,
\end{align}
and $B_{n}$ are the Bernoulli numbers and $\sigma_m(n)$ are divisor sums. We write $g_2=60 G_4$, $g_3=140 G_6$ and suppress the $\tau$ dependence.
\begin{enumerate}
\item We have the following relation to theta functions:
\begin{subequations}
\begin{align}
G_2(\tau)&=-\frac{\vartheta_1'''(0|\tau)}{3\vartheta_1'(0|\tau)}\ , \\
G_4(\tau)&=\frac{\pi^4}{45}(\vartheta_2(\tau)^8-\vartheta_2(\tau)^4\vartheta_3^4+\vartheta_3(\tau)^8) , \\
G_6(\tau)&=\frac{\pi^6}{945}(2\vartheta_2(\tau)^{12}-3\vartheta_2(\tau)^8\vartheta_3(\tau)^4-\vartheta_2(\tau)^4\vartheta_3(\tau)^8+2\vartheta_3(\tau)^{12})\ .
\end{align}
\end{subequations}
see \cite[page 361]{Bateman:1953}. Here, $\vartheta_\nu(\tau)\equiv \vartheta_\nu(0|\tau)$ are the Jacobi theta functions. We eliminated $\vartheta_4(\tau)$ by using the Jacobi abstruse identity
\be 
\vartheta_3(\tau)^4-\vartheta_4(\tau)^4-\vartheta_2(\tau)^4=0\ .
\ee
\item Zeros of $G_4$ and $G_6$. Both Eisenstein series have only a single zero. From the valence formula of modular forms one has
\begin{align}
G_4(\mathrm{e}^{\frac{\pi i}{3}})&=0\ ,  &
G_6(i)&=0\ .
\end{align}
\end{enumerate}

\subsection{The Weierstrass \texorpdfstring{$\wp$}{p} function}
We need the following properties of the Weierstrass $\wp$ function:
\begin{enumerate}
\item Differential equation of $\wp(z|\tau)$:
\be 
\wp'(z|\tau)^2=4\wp(z|\tau)^3-g_2 \wp(z|\tau)-g_3 \ . \label{eq:Weierstrass differential equation}
\ee
\item Addition theorem for $\wp(z|\tau)$:
\be 
\wp(z_1+z_2|\tau)=\frac{1}{4}\left(\frac{\wp'(z_1|\tau)-\wp'(z_2|\tau)}{\wp(z_1|\tau)-\wp(z_2|\tau)}\right)^2-\wp(z_1|\tau)-\wp(z_2|\tau)\ .
\ee
\item Duplication formula for $\wp(z|\tau)$:
\be 
\wp(2z|\tau)=-2\wp(z|\tau)+\frac{1}{4} \left(\frac{\wp''(z|\tau)}{\wp'(z|\tau)}\right)^2 \label{eq:Weierstrass duplication formula}
\ee
\end{enumerate}

\bibliographystyle{JHEP}
\bibliography{bib}

\end{document}